\documentclass[aps,prl,reprint,10pt,twocolumn,footinbib,longbibliography]{revtex4-2}

\usepackage{amsmath,amsthm,amsfonts,amssymb,bm,mathrsfs}
\usepackage{mdframed,float,graphicx,array,MnSymbol,caption,subcaption,placeins}
\usepackage[usenames,dvipsnames]{xcolor}
\definecolor{OceanBlue}{rgb}{0,0.35,0.7} 
\usepackage{hyperref}
\hypersetup{colorlinks=true,allcolors=OceanBlue}
\newtheorem{thm}{Theorem}[section] 
\newtheorem{prop}{Proposition}[section] 

\theoremstyle{definition}

\newcommand{\pars}[1]{\left(#1\right)} 
\newcommand{\bracs}[1]{\left[#1\right]} 
\newcommand{\braces}[1]{\left\{#1\right\}} 

\newcommand{\mr}{\mathrm} 
\newcommand{\mb}{\mathbf} 
\newcommand {\cl}{\mathcal} 
\newcommand {\tsf} [1]{\textsf{#1}} 
\newcommand{\vectsym}{\boldsymbol}  
\newcommand{\C}{\mathbb{C}}     
\DeclareMathOperator{\Tr}{Tr\,}            
\DeclareMathOperator{\tr}{tr\,}            

\newcommand*\diff{\mathop{}\!\mathrm{d}}
\DeclareMathOperator{\trans}{\textsf{T}} 

\newcommand{\balpha}{\vectsym{\alpha}}  
\newcommand{\bxi}{\vectsym{\xi}} 
\newcommand{\bl}{\mb{l}} 
\newcommand{\bn}{\mb{n}} 
\newcommand{\br}{\mb{r}} 
\newcommand{\frke}{\mathfrak{e}} 

\newcommand {\ket}[1] {\left|{#1}\right\rangle}
\newcommand {\bra}[1] {\langle{#1}|}
\newcommand{\braket}[2]{\langle{#1}|{#2}\rangle}

\newcommand{\varket}[1] {\left|{#1}\right\rrangle}
\newcommand{\varbra}[1] {\left\llangle #1\right|}

\newcommand {\kets} [1] {\left|#1\right\rangle_{S}}
\newcommand {\bras} [1] {\left\langle#1\right|\hspace{0mm}_{S}}

\newcommand {\keti} [1] {\left|#1\right\rangle_{I}}

\newcommand {\ketis} [1] {\left|#1\right\rangle_{IS}}

\newcommand{\ahat} {\hat{a}}

\newcommand{\abs}[1]{\left | #1 \right |}   
\newcommand {\norm} [1] {\left \| #1 \right \|} 
\newcommand{\mean}[1]{\left\langle #1 \right\rangle}  

\begin{document}
\title{Quantum limits of covert target detection}
\author{Guo Yao Tham$^{1}$}
\email{ tham0157@e.ntu.edu.sg}
\author{Ranjith Nair$^{1}$}
\email{ranjith.nair@ntu.edu.sg}
\author{Mile Gu$^{1,2,3}$}
\email{gumile@ntu.edu.sg}
\affiliation{$^1$Nanyang Quantum Hub, School of Physical and Mathematical Sciences,  \\
 Nanyang Technological University, 21 Nanyang Link, Singapore 637371 \\
$^2$Centre for Quantum Technologies, National University of Singapore, 3 Science Drive 2, Singapore 117543\\
$^3$MajuLab, CNRS-UNS-NUS-NTU International Joint Research Unit, UMI 3654, 117543, Singapore.}
\begin{abstract} In covert target detection, Alice attempts to send optical or microwave probes to determine the presence or absence of a weakly-reflecting target embedded in thermal background radiation within a target region, while striving to remain undetected by an adversary, Willie, who is co-located with the target and collects all light that does not return to Alice. We formulate this problem in a realistic setting and derive quantum-mechanical limits on Alice's error probability performance in entanglement-assisted target detection for any fixed level of her detectability by Willie. We demonstrate how Alice can approach this performance limit using two-mode squeezed vacuum probes in the regime of small to moderate background brightness, and how such protocols can outperform any conventional approach using Gaussian-distributed coherent states. In addition, we derive a universal performance bound for non-adversarial quantum illumination without requiring the passive-signature assumption.\end{abstract}


\date{\today}
\maketitle

Alice wishes to interrogate a distant region embedded in a thermal background for the presence or absence of a target adversary by probing it with a microwave or optical beam and monitoring the resulting reflections. Meanwhile, the adversary, Willie, monitors his thermal background for statistical deviations from thermal noise, aiming to detect if Alice is actively probing him. In this cat-and-mouse game, \emph{how can Alice maximize her probability of correctly detecting Willie while minimizing Willie's chances of knowing he is being probed}? This question falls under the domain of \emph{covert sensing} and naturally arises in the adversarial arms race between radar and radar detectors~\cite{pace2004detecting}. 

Alice thus faces a trade-off between performance and covertness. Sending a probe with greater energy relative to the background can better sense Willie but also risks being detected. A better idea would be to prepare coherent state probes $\{\ket{\alpha}\}$ with field amplitude $\alpha$ Gaussianly distributed in phase space around the origin to perfectly mimic the statistics of the thermal background, which allows some chance of detecting Willie while being perfectly covert. However, what is the ultimate limit of Alice's performance, and would moving towards this limit be facilitated by employing nonclassical light, as is well-known for quantum illumination  (henceforth abbreviated as QI)  in non-adversarial settings~\cite{TEG+08,PBG+18,PVS+20,Sha20,TB-WK21,STG+22,KFS+23,GTB24}? Indeed, several other covert protocols have been introduced in the quantum continuous-variable setting \cite{BGP+15,BGG+20,DCYP+21,BGD+17,GBD+19,HSG+22}.

To answer our question, we suppose that Alice wishes to remain $\epsilon$-covert, i.e., that the probability of Willie detecting her is at most $1/2 +\epsilon$. We then ask: What is Alice's minimal error probability for detecting Willie? We obtain a closed-form lower bound on this error probability as a function of $\epsilon$, the number of available optical modes $M$, and levels of loss and noise in the system. We show that two-mode squeezed vacuum (TMSV) probes can approach this limit in certain regimes. Comparing TMSV performance with that of the aforementioned Gaussian-distributed coherent state (GCS) probes, our results show that TMSV probes enable a reduction in error probability that scales exponentially with $M$.  

To ensure our bounds apply to all adversaries, we assume Willie can detect any statistical deviation from the thermal background noise -- including those resulting from using multi-mode vacuum probes. This involves dropping the commonly adopted ‘no passive signature’ (NPS) assumption in quantum illumination~\cite{TEG+08,PBG+18,PVS+20,Sha20,TB-WK21,STG+22,KFS+23,GTB24}, a mathematically expedient but nonphysical approximation whereby the background temperature depends in a fine-tuned manner on whether the target is present or absent so as to ensure Alice cannot detect Willie using a vacuum probe. The new techniques we develop for this more mathematically complex setting have immediate relevance to general QI, supporting recent interest in dropping the NPS assumption therein~\cite{JLI+21,LJJ+22,SPC23,Vol24}. Indeed, our work provides the first universal performance limit for QI without the NPS approximation as a natural byproduct.

 \begin{figure*}[t]
    \includegraphics[scale=0.1]{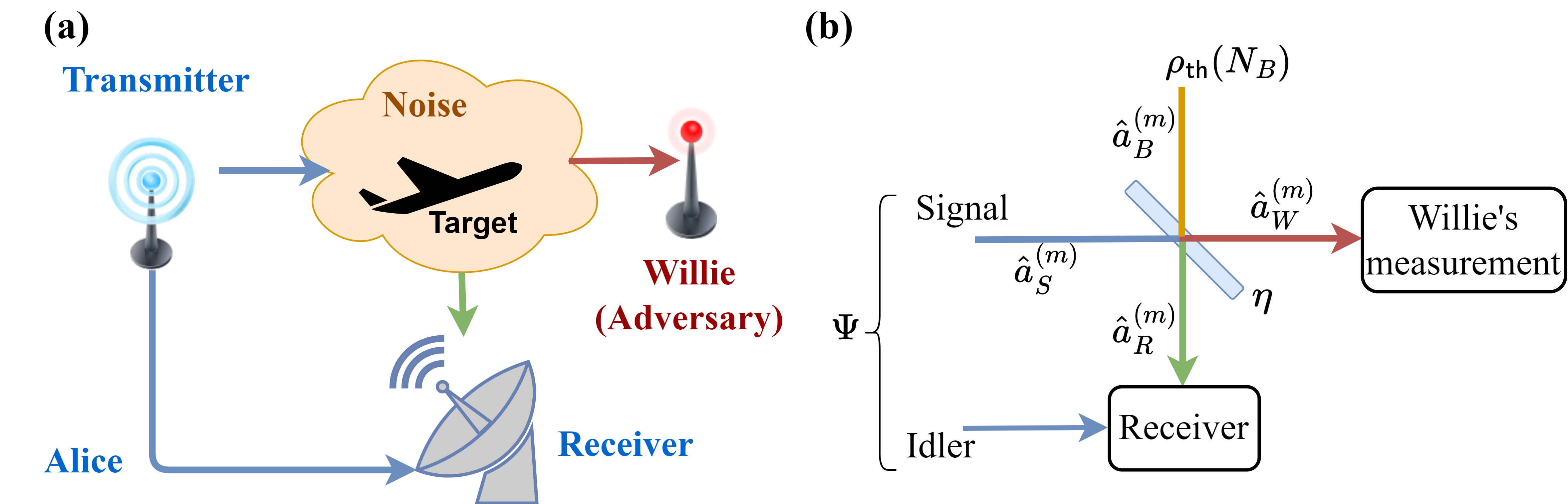}
    \caption{ (a) In covert target detection, Alice (A) attempts to detect the presence of the adversary Willie (W) using an ancilla-entangled probe while remaining undetected herself. In the beam-splitter model (b), Alice prepares a joint state $\Psi$ with $M$ signal ($S$) and idler modes ($I$). Each signal mode $\ahat_S^{(m)}$ is either replaced by a background mode $\ahat_B^{(m)}$ when Willie is absent, or mixed with the background at a beam splitter with reflectance  $\eta\ll 1$ representing the target. Alice makes an optimal measurement on the return modes $\ahat_R^{(m)}$, $m = 1,\ldots M$ along with the idler system.  Willie, when present, makes an optimal measurement on all his modes $\ahat_W^{(m)}$}.
    \label{fig:setup}
\end{figure*}

\indent\emph{Problem Setup---} Target detection is illustrated in Fig.~\ref{fig:setup}. Alice (A) wishes to detect the absence ($h=0$) or presence ($h=1$) of a weakly reflecting target (the two cases being assumed to be equally likely for simplicity) with reflectivity $\eta \ll 1$ \footnote{$\eta$ represents the residual reflectivity in the entire trip from transmitter to receiver and thus includes diffraction losses as well as the reflectivity of the target object itself.}. The target is immersed in a thermal background such that each background mode is in a thermal state $\rho_{\tsf{th}}\pars{N_B}=\sum_{n=0}^\infty N_B^n/(N_B+1)^{n+1}\ket{n}\bra{n}$ with average photon number $N_B$. Alice controls both the transmitter and receiver \footnote{If the radar configuration is bistatic, we assume that the idler modes can be transported losslessly to the receiver's location.}, and can prepare $M$ signal modes. Thus, any covert sensing protocol involves preparing some incident probe state
\begin{align} \label{probe}
\ketis{\psi} = \sum_{\mb{n}} \sqrt{p}_{\mb{n}}\ket{\chi_{\mb n}}_I \ket{\mb n}_S,
\end{align}
where $\ket{\mb n}_S = \ket{n_1}_{1}\ket{n_2}_{2}\cdots \ket{n_M}_{M}$ is an $M$-mode number state of the \emph{signal} ($S$) system, $\{\ket{\chi_{\mb n}}_I\}$ are normalized (not necessarily orthogonal) states of an \emph{idler} ($I$) system, and $p_{\mb n}$ is the probability mass function (pmf) of $\mb{n}$. The signal modes are sent to probe the target region while the idler is held losslessly.  In the return (R) modes, Alice obtains an $h$-dependent return-idler state $\rho^{(h)}_{IR}$  that she measures to make a guess $h_{\mathrm{est}}$ for the value of $h$. Alice's performance is given by the \emph{error probability} $P_\frke^A$, i.e., the probability that $h_{\mathrm{est}} \neq h$. 

The adversary, Willie (W), is situated at the target's location. He is constrained only by the laws of physics and can have prior knowledge of which probe $\Psi = \ket{\psi}\bra{\psi}_{IS}$ Alice plans to use. Thus any statistical deviation of the state intercepted by Willie from the $M$-mode thermal background $\rho_{\tsf{th}}\pars{N_B}^{\otimes M}$ allows him to achieve an error probability $P_\frke^W$ that is better than guessing randomly. Alice's probe state is said to be $\epsilon$-covert (for discussion of the case of unequal prior probabilities for Willie, see Appendix C) if 
\begin{align} \label{ecovertness}
P_\frke^W \geq 1/2 -\epsilon.
\end{align}
We then ask: What is Alice's minimal error probability $P_\frke^A$ (as a function of $M$) when optimized over $\epsilon$-covert probes?

We model the weakly reflecting object by a beam splitter with reflectance $\eta \ll 1$ \cite{Note1}. Let  $\ahat^{(m)}_S$ and $\ahat_B^{(m)}$ be annihilation operators of the corresponding signal and background modes (see Fig.~\ref{fig:setup}). Then 
\begin{align}
    \ahat_R^{(m)}=\sqrt{\eta^{(h)}}\,\ahat_S^{(m)}+\sqrt{1-\eta^{(h)}}\,\ahat_B^{(m)},\label{aR}
\end{align}
represents the annihilation operator of the $m^{\mathrm{th}}$ mode returning to Alice, where $\eta^{(0)} =0$ and $\eta^{(1)} = \eta$. When  {Willie} is present ($h=1$), he receives the modes $\ahat_W^{(m)}$ for each $m = 1, \ldots, M$ from the other output of the beam splitter so that
\begin{align}
    \ahat_W^{(m)}=\sqrt{1-\eta}\,\ahat_S^{(m)} - \sqrt{\eta}\,\ahat_B^{(m)}.\label{aW}
\end{align}
Thus Alice faces the hypothesis test
\begin{equation} \label{Aliceetest}
\begin{aligned}
\mr{H}_0: \; \rho_0 &= \pars{\Tr_S \Psi} \otimes {\rho_\tsf{th}\pars{N_B}}^{\otimes M},\\
\mr{H}_1: \; \rho_1 &= \pars{\mr{id}_I \otimes \cl{L}_{\eta, N_B}^{\otimes M}}\Psi,
\end{aligned}
\end{equation}
where $\cl{L}_{\kappa, N}$ denotes a thermal loss (or noisy attenuator) channel of transmittance $\kappa$ and excess noise $N$ \cite{Ser17qcv}. Meanwhile, Willie faces his own hypothesis test
\begin{equation} \label{Willietest}
\begin{aligned}
\mr{H}_0': \; \sigma_0 &= {\rho_\tsf{th}\pars{N_B}}^{\otimes M},\\
\mr{H}_1': \; \sigma_1 &=  \cl{L}_{1-\eta, N_B}^{\otimes M}\pars{\Tr_I \Psi}.
\end{aligned}
\end{equation}
to decide whether Alice has sent a probe ($\mr{H}_1'$) or not ($\mr{H}_0'$). 

Assuming that both parties make optimal quantum measurements and that their hypotheses are equally likely, their resulting average error probabilities are given by the Helstrom formula~\cite{Hel76}:
\begin{align} 
    P_\frke^A&=1/2-\norm{\rho_0-\rho_1}_1/4 \leq \inf_{0\leq s\leq 1} \Tr \rho_0^s \rho_1^{1-s}/2,\\ 
    P_\frke^W&=1/2-\norm{\sigma_0-\sigma_1}_1/4 \leq \inf_{0\leq s\leq 1} \Tr \sigma_0^s \sigma_1^{1-s}/2,
\end{align}
where $\norm{X}_1:=\Tr\sqrt{X^\dag X}$ is the trace norm. We have also indicated the \emph{quantum Chernoff bound} \cite{ACM+07} (QCB) that is an exponentially tight upper bound on the average error probability.


The above framework deviates in several significant ways from previous studies of covert target detection~\cite{TBG+21}. Firstly, our notion of $\epsilon$-covertness defined by way of Willie's error probability has clear operational significance.
Previous formulations use the relative entropy \cite{TBG+21}, which provides no upper bound on the error probability~\cite{Aud14} and cannot therefore be used to derive a performance limit in our setting. Secondly, our framework fixes the background brightness at $N_B$ regardless of whether a target is present or absent. In contrast, prior work makes the NPS assumption which fine-tunes background brightness from its nominal value of $N_B$ to $N_B/\pars{1-\eta}$ when the target is present \footnote{Specifically, under the NPS assumption, the channel between $S$ and $R$ is taken to be $\cl{L}_{\eta,N_B/\pars{1-\eta}}$ in each mode and that between $S$ and $W$ is $\cl{L}_{1-\eta,N_B/\pars{1-\eta}}$ in each mode when Willie is present, while  $\sigma_0$ is taken to be $\cl{L}_{1-\eta,N_B/\pars{1-\eta}}^{\otimes M}\pars{\ket{0}\bras{0}^{\otimes M}} = \rho_{\tsf{th}}\pars{\eta N_B/\pars{1-\eta}}^{\otimes M}$.}. Under this assumption,  Willie's null hypothesis is to receive multi-mode vacuum states from Alice~\cite{TBG+21}. In our setting, Willie, being bathed in thermal radiation from all directions, receives a thermal state with the same brightness as the background when Alice is absent. Thus, Willie's null hypothesis is based on multi-mode probes that are statistically identical to the thermal background. 

Dropping the NPS assumption induces new qualitative behaviour in both non-adversarial and covert QI. Alice's performance then explicitly depends on the number $M$ of signal modes, e.g. the available time-bandwidth product for temporal modes. 
This is in line with findings for other quantum sensing and discrimination problems for which the mode number is an important resource aiding the performance even when vacuum probes are used \cite{NTG22,JDC22,SZ23,NG23arxiv}. For $\epsilon$-covert illumination, we can quantify the performance of a target detection protocol by defining its Chernoff exponent $\chi:= -\lim_{M \rightarrow \infty}\frac{1}{M}\ln P_\frke^A$. A difference $\Delta$ in the exponents of two probes implies a ratio of $e^{-\Delta M}$ between their error probabilities $P_\frke^A$, which scales exponentially in $M$.

\indent\emph{Illumination with Passive Signature--} We first derive analytical bounds on Alice's performance in the non-adversarial setting, i.e., for standard QI but without the NPS approximation. In Appendix~B, we show a more general result: For $\tsf{b} \in \braces{0,1}$, let states $\rho_\tsf{b}:= \pars{\mr{id}_I \otimes \cl{L}_{\kappa_\tsf{b},N_{\tsf{b}}}^{\otimes M}}\Psi$ be the respective output states of any two thermal loss channels $\cl{L}_{\kappa_0,N_0}$ and $\cl{L}_{\kappa_1,N_1}$ in response to the input $\Psi$ of Eq.~\eqref{probe}. Then the output fidelity $F:= \Tr \sqrt{\sqrt{\rho_0} \rho_1 \sqrt{\rho_0}}$ satisfies
\begin{align} \label{fidlb}
F \geq \nu^M \sum_{n=0}^\infty p_n \bracs{\nu\sqrt{\widetilde{\kappa}_0 \widetilde{\kappa}_1} + \sqrt{\pars{1-\widetilde{\kappa}_0}\pars{1-\widetilde{\kappa}_1}}}^n,
\end{align}
where $\nu = \pars{\sqrt{G_0G_1} - \sqrt{\pars{G_0-1}\pars{G_1-1}}}^{-1}$ and $\widetilde{\kappa}_\tsf{b} = \kappa_\tsf{b}/G_\tsf{b}$ for $G_\tsf{b} = \pars{1-\kappa_\tsf{b}}N_\tsf{b} +1$.
For target detection, we set $\kappa_0 = 0, \kappa_1 = \eta, N_0 = N_1 = N_B$ and use the Fuchs-van de Graaf inequality \cite{FvdG99} to conclude that
\begin{align} \label{PeAlb}
P_\frke^A &\geq \frac{1}{2}\bracs{1-\sqrt{1-\nu^{2M}\bracs{\sum_{n=0}^\infty p_n \pars{1-\gamma_{\eta,N_B}}^{\frac{n}{2}}}^2}}\nonumber\\
&\geq \frac{1}{2}\bracs{1-\sqrt{1-\nu^{2M}\pars{1-\gamma_{\eta,N_B}}^{\cl{N}_S} }},
\end{align}
where $\gamma_{\eta,N_B} = \frac{\eta}{(1-\eta)N_B+1}$, $\cl{N}_S := \sum_{n=0}^\infty n p_n$ is the total signal energy, and we have used Jensen's inequality in the last step. The above result gives the ultimate limit of Alice's performance in QI with the PS assumption. It contrasts with the ultimate quantum limits of NPS QI derived in ref.~\cite{NG20} (see Eqs.~(12)-(13) therein), which do not include the $M$-dependent factor $\nu^{2M}$ characteristic of the passive signature.


\indent\emph{Necessary condition for $\epsilon$-covertness---} 
To incorporate the covertness constraint, we now formulate a necessary condition for $\epsilon$-covertness. 
Suppose that Alice transmits the probe $\Psi$ of Eq.~\eqref{probe} with signal energy $\cl{N}_S$. 
The Fuchs-van de Graaf inequality $P_\frke^W \leq F\pars{\sigma_0,\sigma_1}/2$ that relates the trace distance to the fidelity \cite{FvdG99} between Willie's hypothesis states \eqref{Willietest} implies that $F\pars{\sigma_0,\sigma_1} \geq 1 -2\epsilon$ is a necessary condition for $\epsilon-$covertness. In Appendix~C, we use this to show that
\begin{align} \label{ecovcondition}
    \sum_{n=0}^\infty\sqrt{q_n}\sqrt{\binom{n+M-1}{n}\frac{N_B^n}{(N_B+1)^{n+M}}}\geq 1-2\epsilon
\end{align}
is a necessary condition for $\epsilon-$covertness, where $q_\bn = \bra{\bn}\sigma_1\ket{\bn}_W$ and $\braces{q_n = \sum_{\bn: n_1 + \cdots + n_M} q_\bn }_{n=0}^\infty$ is the pmf of the total photon number seen by Willie under $\mr{H}'_1$.

\begin{figure} 
         \includegraphics[trim=0mm 60mm 0mm 65mm, clip=true,scale=0.4]{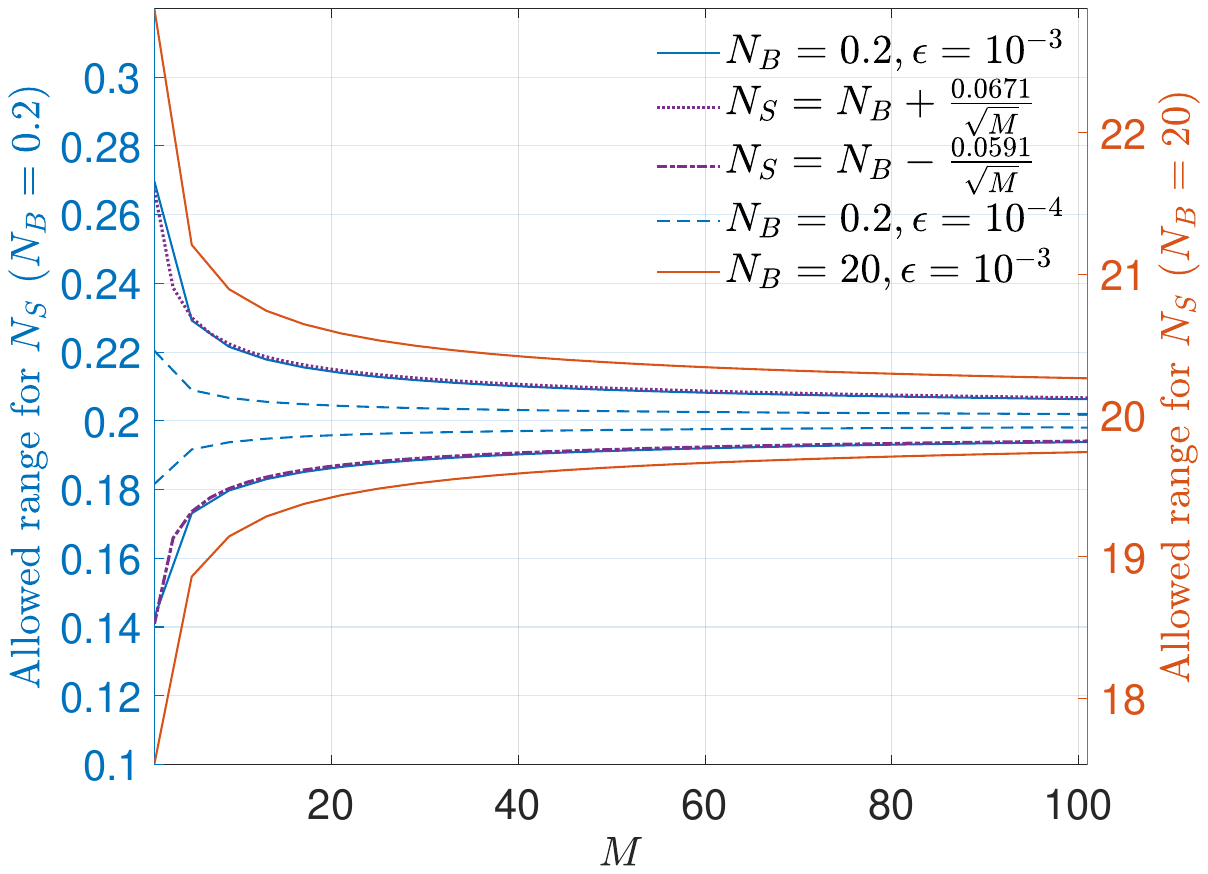}  
        \caption{The maximum and minimum allowed per-mode energy $N_S$ for an $\epsilon$-covert probe according to Eq.~\eqref{ecovcondition} as a function of $M$ with $N_B=0.2$  and $\epsilon = 10^{-3}$ (solid blue) and $\epsilon=10^{-4}$ (dashed blue). Curve fitting produced the estimates $ N_B +0.0671 /\sqrt{M}$ and $  N_B - 0.0591/\sqrt{M}$ for the maximum and minimum energy curves for $\epsilon=10^{-3}$.   The allowed range of $N_S$ for $N_B=20,\epsilon=10^{-3}$ is also shown (red). $\eta = 0.01$ for all curves. }
        \label{fig:energy_limits}
\end{figure}

\indent\emph{Signal Energy Constraints ---} The above condition then implies bounds on the allowed signal energies. Clearly, for $\epsilon = 0$, an $M$-mode quantum or classical probe with perfect covertness must have signal energy  $\cl{N}_S=MN_B$ to match perfectly with Willie's thermal background. For $\epsilon>0$, we use Lagrange multipliers to extremize the average energy $\sum_{n=0}^\infty nq_n$ of $\sigma_1$ under the constraint of Eq.~\eqref{ecovcondition}, which then constrains the probe energy $\cl{N}_S  = \bracs{\sum_{n=0}nq_n - \eta N_B}/(1-\eta)$ (cf. Eq.~\eqref{aW}). We find (Fig.~\ref{fig:energy_limits}) that the probe energy must lie in a bounded region around $MN_B$ that gets smaller as $\epsilon$ is reduced -- see Appendix~D for details. Curve fitting indicates that the per-mode probe energy $N_S := \cl{N}_S/M$ varies between $\sim N_B  \pm A_{\pm}/\sqrt{M}$, where  $A_+$ and $A_-$ depend only on $\eta,N_B$ and $\epsilon$ (See Fig.~\ref{fig:energy_limits}). To maintain covertness, Alice's per-mode probe must look progressively more similar to the thermal background as we increase $M$. In significant departure from standard QI, it no longer makes operational sense to consider the scaling of Alice's performance with signal energy $\cl{N}_S$. Instead, the key resource is the number of available optical modes $M$.

\indent\emph{Fundamental limits under $\epsilon$-covertness---} The thermal loss channel $\cl{L}_{1-\eta, N_B}$ connecting the modes in $S$ to those in $W$ (cf. Eq.~\eqref{Willietest}) admits the decomposition
\begin{align} \label{channeldecomp}
\cl{L}_{1-\eta,N_B} = \cl{A}_G \circ \cl{L}_{(1-\eta)/G}
\end{align}
into a quantum-limited amplifier $\cl{A}_G$ of gain $G=\eta N_B +1$ and a pure-loss channel $\cl{L}_{(1-\eta)/G}$ of transmittance $(1-\eta)/G$ \cite{CGH06}.  The right-hand side of the bound of Eq.~\eqref{PeAlb} is expressed in terms of the \emph{probability generating function} (pgf) $\cl{P}_S\pars{\xi}$ of the total photon number in $S$, defined as 
$\cl{P}_S\pars{\xi} := \sum_{n=0}^\infty p_n \,\xi^n$
evaluated at $\xi = \sqrt{1 - \gamma_{\eta,N_B}}$. Haus has developed relations connecting the input and output photon number pgfs of these single-mode quantum-limited channels \cite{Hau00enqom}. In Appendix~E, we use the decomposition \eqref{channeldecomp} to extend these to multimode thermal loss channels and find the one-to-one mapping between the photon number pgf of the probe and the pgf $\cl{P}_W\pars{\xi}:= \sum_{n=0}^\infty q_n \xi^n$  of the total photon number in Willie's modes under $\mr{H}'_1$. By connecting $\cl{P}_W\pars{\xi}$  to the covertness condition of Eq.~\eqref{ecovcondition}, we show in Appendix~F that 
\begin{align} \label{PeAlbcov}
 P_\frke^A \geq \frac{1- \sqrt{1- \pars{1-2\epsilon}^4 f^{2M}_{\eta,N_B}}}{2},
\end{align}
where $f_{\eta,N_B} = \nu(N_B+1-\frac{N_B}{x})[\eta N_B(1-x) + 1]$, $x=1 - \frac{\Theta}{\eta[1+N_B(1-\Theta)]}$ and $\Theta=\frac{\sqrt{(1-\eta)(N_B+1)}}{\sqrt{1 + (1-\eta)N_B}}$. While this equation looks complex, it provides a universal, analytical and probe-independent lower bound for Alice's error probability for any desired covertness level $\epsilon$.

\begin{figure}
         \includegraphics[trim=10mm 60mm 5mm 65mm, clip=true,scale=0.4]{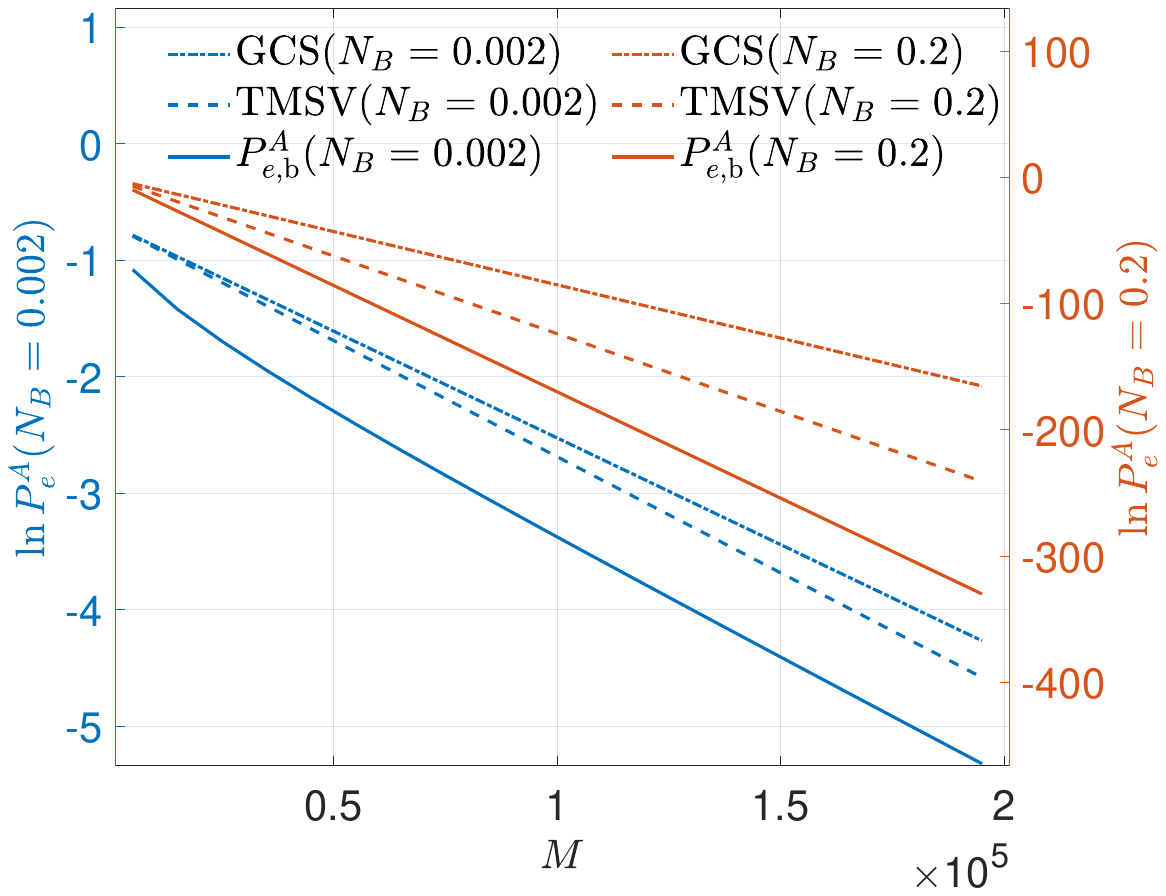}  
          \caption{The lower bound  Eq.~\eqref{PeAlbcov} (solid) on Alice's error probability is compared to that of $\epsilon$-covert TMSV (dashed) and GCS probes (	dash-dotted line) for $N_B =0.2$ (blue) and $N_B = 0.002$ (red). $\epsilon=10^{-3}$ for both. For large $M$, the ratio of the error exponents predicted by the bound \eqref{PeAlbcov}  and of TMSV probes are $1.37$ (for  $N_B = 0.2$)  and $1.16$ (for $N_B=0.002$) respectively.} \label{fig:PeAcomp}
\end{figure}

Figure~\ref{fig:PeAcomp} compares the lower bound of Eq.~\eqref{PeAlbcov} to quantum TSMV and classical GCS probes. For each $M$, we consider the $M$-mode independently and identically distributed (iid) TMSV state with per-mode signal energy $N_S$ chosen to be the maximum allowable by the covertness constraint (see Appendix~F for details). When limited to classical probes, Alice can generate Gaussian-distributed coherent state (GCS) probes -- coherent states in each signal mode with amplitude $\alpha \in \mathbb{C}$ chosen according to a product circular Gaussian distribution $P\pars{\alpha} = \frac{1}{\pars{\pi N_S}}e^{-\abs{\alpha}^2/N_S}$ with identical per-mode energy as for the TMSV probe (Note that, in this case, Alice's measurement can be conditioned on her knowledge of the amplitude transmitted in each of the $M$ shots). For $N_B=0.2$, the large-$M$ error exponent achieved by TMSV probes was about a factor of 1.37 lower than that of the bound, with the discrepancy becoming smaller for smaller $N_B$, along with the gap between the GCS and TMSV exponents.

\begin{figure}[tbh] 
            \includegraphics[trim=0mm 60mm 14mm 63mm, clip=true,scale=0.42]{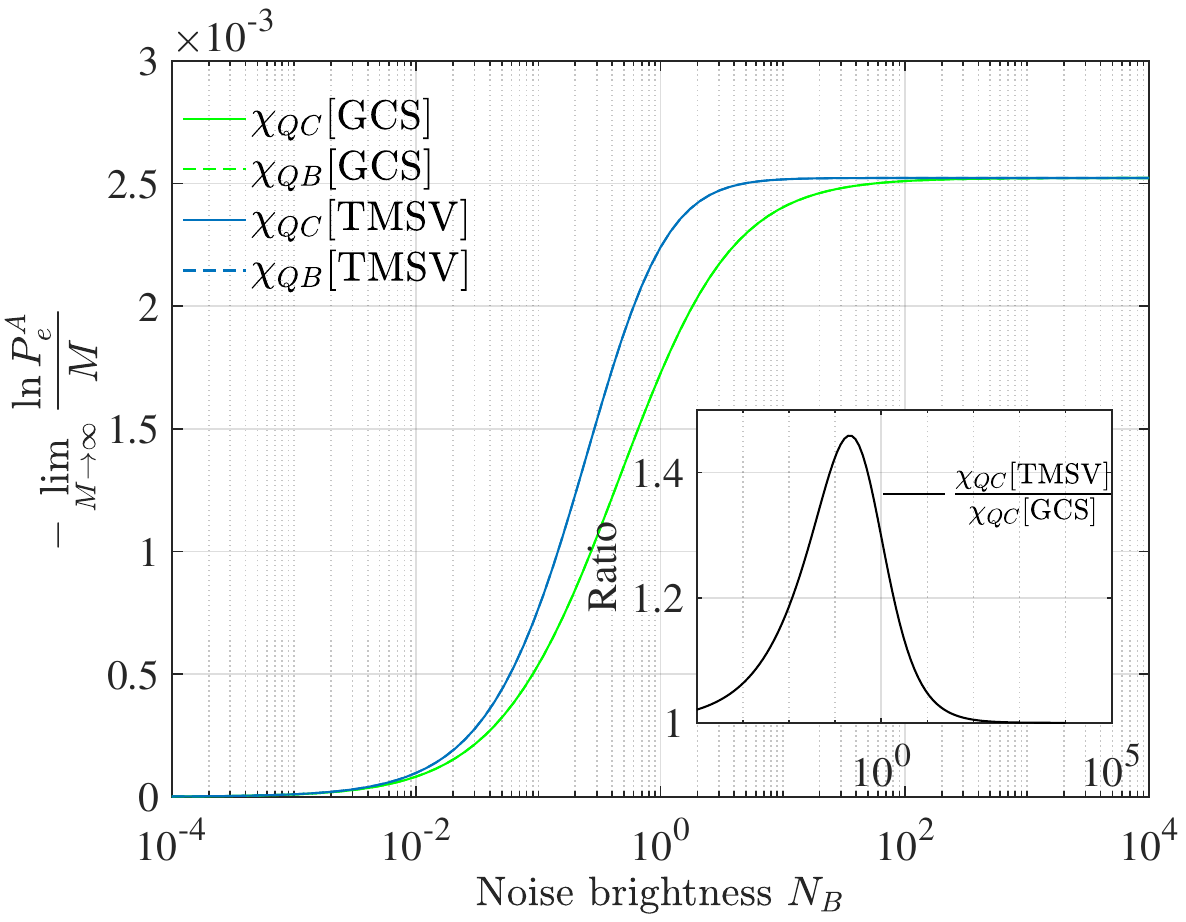}
     \caption{The quantum Chernoff (QC) (solid) and {(closely overlapping)} quantum Bhattacharyya (QB) (dashed) error probability exponents as a function of the noise brightness $N_B$ with $\eta=0.01$ for perfectly covert TMSV (blue) and GCS (green) probes. The inset shows the ratio of the exponents for the TMSV and GCS probes, which is maximized for $N_B \simeq 0.2$.}\label{fig:qcb_perfect}
\end{figure}

\indent\emph{Perfect covertness --} Consider the special case of perfect covertness, i.e., $\epsilon=0$. Among quantum probes, the $M$-mode TMSV state with signal energy $N_S = N_B$ is the only viable pure state (modulo unitary transformations on the idler) quantum probe. Among classical probes, only the iid GCS probes with $N_S = N_B$ are possible. 
To compare their target detection performance, we calculate the quantum Chernoff exponents $\chi$ (Appendix~A). These are very close to the quantum Bhattacharyya $(s=1/2)$ exponents, which have the approximate forms:
\begin{equation}
\begin{aligned}
 \chi_{\mr{TMSV}} &\simeq \ln \bracs{1 - \frac{\eta}{4}\bracs{1 - \frac{1}{(2N_B + 1)^2}}},\\
\chi_{\mr{GCS}} &\simeq  \ln \bracs{1- 2\eta N_B \pars{N_B -\sqrt{N_B(N_B+1)} + \frac{1}{2}}}.
\end{aligned}
\end{equation}
The results are compared in Fig.~\ref{fig:qcb_perfect}. Observe $\chi_{\mr{TMSV}} \geq \chi_{\mr{GCS}}$ everywhere with a maximum ratio of $\simeq 1.45$ around $N_B \simeq 0.2$. For example, for $N_B = 0.575$ and $M = 10^6$ (current technology already reaches $M \sim 10^6$ \cite{Sha20}), TMSV probes  offer a $\sim 10^{237}$-fold reduction in error probability.

\indent\emph{Discussion---} We introduced an operational framework for covert quantum target detection: Alice wishes to minimize her error probability in detecting an adversary Willie in thermal noise, subject to an $\epsilon$-covertness constraint to ensure Willie's error probability in detecting her remains above $1/2 - \epsilon$. Our lower bound \eqref{PeAlbcov} is a fundamental limit on Alice's performance. {As Fig.~\ref{fig:PeAcomp} indicates, the TMSV error exponent is close to that of our lower bound in the optical regime of $N_B \ll 1$, with the maximum advantage over the Gaussian-distributed coherent states being obtained for $N_B \simeq 0.2$. In certain regimes, the achievable quantum advantage --- measured by the factor by which entangled probes can reduce error probabilities -- scales exponentially with the number of optimal modes $M$. Our framework assumes Willie can optimally detect any deviation from the thermal background noise of brightness $N_B$ and includes  the first universal performance limits for standard QI with passive signature.

Several interesting open questions remain. Towards practical realization, a natural question is how the quantum advantage in covert target detection can be harnessed using linear-optics-based measurements similar to standard QI ~\cite{SZS+24}. Other interesting problems are to verify the conjectures that the GCS scheme studied here gives the lowest $P_\frke^A$ among all \emph{classical} schemes under $\epsilon$-covertness (this is clearly the case for $\epsilon=0$), and if the TMSV probes are indeed optimal quantum ones for arbitrary $N_B$ (cf. the numerical study \cite{BCT+21} for standard QI). Meanwhile, considering scenarios where Willie is further constrained by practical constraints (e.g. when he cannot collect all the modes that do not return to Alice after reflection at the target) is of great interest to explore further use cases of QI. Our analytical bounds for non-adversarial QI without the NPS approximation allow us to better understand the regime of validity of this approximation in line with recent research ~\cite{JLI+21,LJJ+22,SPC23,Vol24}. 

Beyond target detection, suitable modification of Willie's hypothesis test and Alice's performance metric could enable applying our techniques to other sensing problems where the no passive signature approximation matters. In particular, as illustrated here, our fidelity-based approach yields unconditional energy bounds on $\epsilon$-covert probes without requiring assumptions on, e.g., their photon number variance ~\cite{BGD+17,GBD+19}. This could then enable explicit $M$-dependent performance bounds for protocols such as covert phase and transmittance sensing \cite{BGD+17,GBD+19,HSG+22}. Finally, our fidelity bounds of  Eqs.~\eqref{fidlb} on the outputs of thermal loss channels should be useful for obtaining fundamental performance bounds for schemes such as quantum reading \cite{Pir11}, pattern recognition  \cite{BZP20} and channel position finding \cite{ZP20,PBZ+21}.

}

\indent\emph{Acknowledgements---} This work is supported by the Singapore Ministry of Education Tier 2 Grant No. T2EP50221-0014, the Singapore Ministry of Education Tier1 Grants RG77/22 and RT4/23, the Agency for Science, Technology and Research (A*STAR) under its QEP2.0 programme (NRF2021-QEP2-02-P06), and FQxI under grant no ~FQXi-RFP-IPW-1903 (``Are quantum agents more energetically efficient at making predictions?") from the Foundational Questions Institute and Fetzer Franklin Fund (a donor-advised fund of Silicon Valley Community Foundation). Any opinions, findings and conclusions or recommendations expressed in this material are those of the author(s) and do not reflect the views of National Research Foundation or the Ministry of Education, Singapore.

\onecolumngrid
\section{Appendix A: Quantum Chernoff exponents for TMSV and GCS transmitters}
\subsection{Gaussian state specification and the quantum Chernoff bound}
In this section, we describe how the quantum Chernoff exponents for TMSV and GCS probes discussed in the main text were obtained. To this end, we recall the formalism of mean vectors and covariance matrices for multimode states. We define a vector of quadrature field operators $Q=(x_1,\ldots,x_M,p_1,\cdots,p_M)^{\trans}$ of an arbitrary $M$-mode  system whose components are the dimensionless operators 
\begin{align}
\hat{q}_m&=\frac{\hat{a}_m+\hat{a}_m^\dag}{\sqrt{2}},\qquad \hat{p}_m=\frac{-i(\hat{a}_m-\hat{a}_m^\dag)}{\sqrt{2}}.
\end{align}
These quadrature operators satisfy the commutation relations:
\begin{align}
[Q,Q^{\trans}]&=i\Omega;\qquad \Omega=\begin{pmatrix}
0 & 1\\-1 & 0
\end{pmatrix}\otimes\mathbb{1}_M.
\end{align}
For a given state $\rho$, we have the \emph{mean vector}
\begin{align}
\overline{Q} &=\mean{Q}=\Tr(Q\hat{\rho})
\end{align}
and the (Wigner) covariance matrix $\widetilde{V}$, whose $ij$-th entry is
\begin{align}
\widetilde{V}_{ij} &=\frac{1}{2}\mean{\{\Delta Q_i,\Delta Q_j\}}\nonumber\\
&=\frac{1}{2}\mean{\{Q_i,Q_j\}}-\mean{Q_i}\mean{Q_j},
\end{align}
where $\Delta Q_i=Q_i-\mean{Q_i}$ and $\{,\}$ is the anticommutator. 
For Gaussian states such as those under consideration here, the mean vector and covariance matrix fully determine the state. 

We recall the quantum Chernoff bound (QCB) that provides an exponentially tight upper bound on the optimal error probability for Alice's discrimination of her hypotheses $H_0$ and $H_1$ \cite{ACM+07}. For a priori equally likely hypotheses, it reads
\begin{align} \label{qcb}
P_{E}^{A}\bracs{\rho_0,\rho_1}\leq \frac{1}{2}\pars{\inf_{0\leq s\leq 1} \Tr \rho_0^s \rho_1^{1-s}} \equiv \frac{1}{2}\pars{\inf_{0\leq s\leq 1} C_s}.
\end{align}
Note that the QCB is multiplicative in the sense that if $\rho_0 = \otimes_{m=1}^M r_{0,m}$ and $\rho_1 = \otimes_{m=1}^M r_{1,m}$, we have
\begin{align} \label{qcbmult}
P_{E}^{A}\bracs{\rho_0,\rho_1}\leq \frac{1}{2}\pars{\inf_{0\leq s\leq 1} \Tr \rho_0^s \rho_1^{1-s}} = \frac{1}{2} \prod_{m=1}^M\pars{\inf_{0\leq s\leq 1} \Tr r_{0,m}^sr_{1,m}^{1-s}}.
\end{align}
Note that choosing any value of $s$ between $0$ and $1$ provides an upper bound on $P_\frke^A$. The case $s=1/2$ is known as the quantum Bhattacharyya bound (QBB). 
For Gaussian states, we can leverage existing techniques in the literature \cite{BBP15} to compute $C_s$, and thence the QCB, once the means and covariances of $\rho_0$ and $\rho_1$ are known.

\subsection{Gaussian-distributed coherent-state (GCS) transmitter}

As discussed in the main text, a possible classical probe that Alice can transmit consists of a string of $M$ coherent states $\ket{\alpha_1}\otimes \cdots \otimes \ket{\alpha_M} \equiv \ket{\balpha}$ with amplitudes drawn independently from a circular Gaussian distribution
\begin{align} \label{circgauss}
P\pars{\alpha} = \frac{1}{\pi N_T}\exp\pars{-\frac{\abs{\alpha}^2}{N_T}},
\end{align}
where $N_T$ is the average per-mode transmitted energy.
The reduced state seen by Willie in each signal mode is then a thermal state $\tau\pars{N_T}$. Alice, on the other hand, knows the vector $\balpha$ and can incorporate this knowledge in her receiver. Accordingly, her error probability is 
\begin{align}
P_\frke^A[\mr{GCS}] = \int_{\mathbb{C}^M} \diff^{2M}\balpha\, P\pars{\balpha} P^A_E\bracs{\balpha},
\end{align}
where $P^A_E\bracs{\balpha}$ is shorthand for Alice's optimal error probability of discriminating the two possible received states when the transmitter is $\ket{\balpha}$. 

For a single mode in which coherent state $\ket{\alpha} \equiv \ket{\alpha_r + i\alpha_i}$ is transmitted, the mean vector and covariance matrix of the received state of $\ahat_R$ under the two hypotheses can be computed using { Eq.~(3) of the main text} as follows:
\begin{align} 
\mr{H}_0: &\,  \overline{Q} =\begin{bmatrix}
0\\0
\end{bmatrix};\qquad \widetilde{V} =\begin{bmatrix}
N_B + \frac{1}{2} & 0\\0 & N_B + \frac{1}{2}
\end{bmatrix}, \\ 
\mr{H}_1:\;&  \overline{Q} =\begin{bmatrix}
\sqrt{2\eta\alpha_r}\\\sqrt{2\eta\alpha_i}
\end{bmatrix};\qquad  \widetilde{V} =\begin{bmatrix}
\pars{1-\eta}N_B + \frac{1}{2} & 0\\0 & \pars{1-\eta}N_B + \frac{1}{2} 
\end{bmatrix}.   
\end{align}
These correspond to a thermal state and displaced thermal state respectively. The passive signature manifests in the fact that these states have slightly different temperatures.

Using the QCB Eq.~\eqref{qcb}, we can bound Alice's error probability for an $M$-mode GCS transmitter as follows:  
\begin{align}
P_\frke^A[\mr{GCS}] &= \int_{\mathbb{C}^M} \diff^{2M}\balpha \pars{\prod_{m=1}^M P\pars{\alpha_m}} P^A_E\bracs{\balpha} \\
&\leq \frac{1}{2} \int_{\mathbb{C}^M} \diff^{2M}\balpha\prod_{m=1}^M \bracs{ P\pars{\alpha_m} \pars{\inf_{0\leq s \leq 1}C_s\bracs{\alpha_m}}}\\
&= \frac{1}{2} \bracs{\int_{\mathbb{C}} \diff^2 \alpha \,P(\alpha) \pars{\inf_{0\leq s \leq 1}C_s\bracs{\alpha}}}^M,
\end{align}
where we have used the multiplicativity of the QCB from Eq.~\eqref{qcbmult}.
In principle, the optimizing $s$ inside the above integral can depend on $\abs{\alpha}$ { but we found numerically that $s=1/2$ was nearly optimum for all values of $\abs{\alpha}$ and $N_B$ in the  $\eta \leq 0.01$ region.} Accordingly, the  optimal error probability exponent equals the Bhattacharyya error exponent:
\begin{align} \label{beegcs}
\chi_{\mr{GCS}} := -\lim_{M \rightarrow \infty} \frac{\ln\,P_\frke^A[\mr{GCS}]}{M} \simeq - \ln \int_{\mathbb{C}} \diff^2 \alpha \,P(\alpha) C_{1/2}\bracs{\alpha}.
\end{align}
For perfect covertness, we set $N_T = N_B$ in the above and the result is approximated in { Eq.~(14) of the main text}.

\subsection{Two-mode squeezed vacuum (TMSV) transmitter}
Consider using multiple copies of the TMSV state  $\ket{\psi}=\sum_{n=0}^\infty\sqrt{\frac{N_S^n}{\pars{N_S+1}^{n+1}}}\kets{n}\keti{n}$. This state has a zero mean vector and the covariance matrix
\begin{align}
\widetilde{V}_{SI} &=\begin{bmatrix}
S & C\\C & S
\end{bmatrix}\oplus\begin{bmatrix}
S & -C\\-C & S
\end{bmatrix},
\end{align}
where
\begin{align}
S &=N_S+\frac{1}{2},\\
C &= \sqrt{N_S(N_S+1)},
\end{align}
and $X\oplus Y=\begin{bmatrix}
X & 0\\0 & Y
\end{bmatrix}$ is the direct sum of matrices.

{Using Eq.~(3) of the main text,} we can compute the first and second moments of the states corresponding to Alice's two hypotheses. Both states remain zero-mean, and the covariance matrices are

\begin{align}
\mr{H}_0:\,\widetilde{V}_{RI} &=
\begin{bmatrix}
B & 0\\0 & S
\end{bmatrix}\oplus\begin{bmatrix}
B & 0\\0 & S
\end{bmatrix},\\
\mr{H}_1:\, 
\widetilde{V}_{RI} &=\begin{bmatrix}
A & \sqrt{\eta}C\\
\sqrt{\eta}C & S
\end{bmatrix}\oplus\begin{bmatrix}
A & -\sqrt{\eta}C\\
-\sqrt{\eta}C & S
\end{bmatrix},
\end{align}
where
\begin{align}
A &= \eta N_S + (1-\eta)N_B + \frac{1}{2}, \\
B &= N_B + \frac{1}{2}.
\end{align}
As for the coherent-state case, the covariance matrices above can be used to numerically compute the Chernoff bounds $C_s$ using the techniques in Ref.~\cite{BBP15}.

\section{Appendix B: Lower bound for Alice's error probability of distinguishing thermal loss channels}

Among the pure-state probes $\Psi = \ket{\psi}\bra{\psi}_{IS}$ with
\begin{align} \label{transmitter}
\ketis{\psi} &= \sum_{\mb{n}} \sqrt{p_{\mb{n}}} \keti{\chi_{\mb{n}}}\kets{\bn},
\end{align}
with a given photon number pmf in $S$, the \emph{NDS probes} are those satisfying $\braket{\chi_{\bn}}{\chi_{\bn'}} = \delta_{\bn,\bn'}$.
It follows from the phase covariance of thermal loss channels and the results of ref.~\cite{SWA+18} (See Lemma 4 therein with trace distance as the divergence measure) that among all probes with a given $\{p_\bn\}$, NDS transmitters 
 have the smallest $P_\frke^A$. Accordingly, we can focus on them in order to derive a lower bound on $P_\frke^A$. 
For an NDS probe of the form of Eq.~\eqref{transmitter}, the hypotheses to be distinguished by Alice are
\begin{align}
\mr{H}_0: & \rho_0 = \pars{\mr{id}_I \otimes \cl{L}_{0,N_B}^{\otimes M}} \Psi = \pars{\mr{id}_I \otimes \pars{\cl{A}_G \circ \cl{L}_{0}}^{\otimes M}} \Psi. \\
\mr{H}_1: & \rho_1 = \pars{\mr{id}_I \otimes \cl{L}_{\kappa,N_B}^{\otimes M}} \Psi= \pars{\mr{id}_I \otimes \pars{\cl{A}_{\widetilde{G}} \circ \cl{L}_{\widetilde{\kappa}}}^{\otimes M}} \Psi,
\end{align}
where 
\begin{align} 
\widetilde{G} &= (1-\kappa)N_B+1, \\
 \widetilde{\kappa} &= \kappa/\widetilde{G}.
 \end{align}
In this section, we prove the somewhat more general result of Theorem \ref{thm:ampattcascadefidlb}. To this end, we need the following result proved in Ref.~\cite{NG23arxiv} (see Appendix therein).

\begin{prop} \label{prop:ampoutputfidelity}
With $\mb{n} = \pars{n_1,\ldots, n_M}$ indexing the $M$-mode Fock states of $S$, consider the NDS states
\begin{equation} \label{states}
\begin{aligned}
\ketis{\psi} &= \sum_{\mb{n}} \sqrt{r_\mb{n}} \keti{\chi_\mb{n}} \kets{\mb{n}}, \\
\ketis{\psi'} &= \sum_{\mb{n}} \sqrt{s_\mb{n}} \keti{\chi_\mb{n}} \kets{\mb{n}},
\end{aligned}
\end{equation}
where $\left\{r_\mb{n}\right\}$ and  $\left\{s_\mb{n}\right\}$ are arbitrary multimode photon probability distributions and $\left\{\keti{\chi_\mb{n}} \right\}$ is a given orthonormal set of states in $A$.  Suppose that
\begin{equation}
\begin{aligned}
\rho &= \pars{\mr{id}_A \otimes \cl{A}_G^{\otimes M } }\pars{\ket{\psi}\bra{\psi}}, \\
\rho' &= \pars{\mr{id}_A \otimes \cl{A}_{G'}^{\otimes M } }\pars{\ket{\psi'}\bra{\psi'}}.
\end{aligned}
\end{equation}
The fidelity between these  output states is given by
\begin{align} \label{ampoutputfidelity}
F\pars{\rho, \rho'} &= \sum_{\mb{n} \geq \mb{0}}  \sqrt{r_\mb{n}\, s_\mb{n}}\;\nu^{n+M}, 
\end{align}
where $n = \sum_{m=1}^M n_m$ 
and 
\begin{equation} \label{nudef}
\nu = \pars{\sqrt{GG'} - \sqrt{\pars{G-1}\pars{G'-1}}}^{-1}\in (0,1].
\end{equation}
\end{prop}

\begin{thm} \label{thm:ampattcascadefidlb}
Consider two cascaded channels 
\begin{equation}
\begin{aligned}
\cl{C} &:=\pars{\cl{A}_G \circ \cl{L}_\kappa }^{\otimes M}, \\
\cl{C}'&:= \pars{\cl{A}_{G'} \circ \cl{L}_{\kappa'} }^{\otimes M},
\end{aligned}
\end{equation}
 acting on an $M$-mode system S and an NDS probe state
\begin{align} 
\ketis{\psi} = \sum_{\mb{n}} \sqrt{p_\mb{n}} \keti{\chi_\mb{n}} \kets{\mb{n}}
\end{align}
and let $\Psi = \ket{\psi}\bra{\psi}_{IS}$. Let 
\begin{align}
p_n = \sum_{\mb{n}\geq \mb{0}: \tr \mb{n} = n} p_\mb{n}
\end{align}
be the probability distribution of the total photon number in $S$. For the output states
\begin{equation}
\begin{aligned}
\rho &= \pars{\mr{id}_I \otimes \cl{C}}\pars{\Psi}, \\
\rho' &= \pars{\mr{id}_I \otimes \cl{C}'}\pars{\Psi},
\end{aligned}
\end{equation}
we have the lower bound 
\begin{align} \label{ketis}
F\pars{\rho, \rho'} \geq \nu^M \sum_{n=0}^\infty p_n \bracs{\nu\sqrt{\kappa \kappa'} + \sqrt{\pars{1-\kappa}\pars{1-\kappa'}}}^n,
\end{align}
on the fidelity between them, where $\nu$ is given by Eq.~\eqref{nudef}.

\begin{proof}
Consider the states 
\begin{equation}
\begin{aligned}
\sigma &= \pars{\mr{id}_A \otimes \cl{L}_{\kappa}^{\otimes M}}\pars{\Psi}, \\
\sigma' &= \pars{\mr{id}_A \otimes \cl{L}_{\kappa'}^{\otimes M}}\pars{\Psi}.
\end{aligned}
\end{equation}
Let $\kappa = \cos^2 \phi$ and $\kappa' = \cos^2 \phi'$. The states above can be written as a mixture of states corresponding to the `loss pattern' $\bl$ of photons to the environment of the attenuator as \cite{Nai18loss}:
\begin{equation}
\begin{aligned}
\sigma &= \sum_{\mb{l} \geq \mb{0}} \varket{\psi_\mb{l}}\varbra{\psi_\mb{l}} \equiv \sum_{\mb{l} \geq \mb{0}} q_\mb{l}\,\ket{\psi_\mb{l}}\bra{\psi_\mb{l}} \equiv \sum_{\mb{l} \geq \mb{0}} q_\mb{l} \Psi_\mb{l}, \\
\sigma' &= \sum_{\mb{l} \geq \mb{0}} \varket{\psi'_\mb{l}}\varbra{\psi'_\mb{l}} \equiv \sum_{\mb{l} \geq \mb{0}} q'_\mb{l}\,\ket{\psi'_\mb{l}}\bra{\psi'_\mb{l}} \equiv \sum_{\mb{l} \geq \mb{0}} q'_\mb{l} \Psi'_\mb{l}, 
\end{aligned}
\end{equation}
with 
\begin{equation} \label{rdefined}
\begin{aligned}
\varket{\psi_\mb{l}} &=
\bracs{\sin \phi}^{\tr \mb{l}} \sum_{\mb{n} \geq \mb{l}} \sqrt{p_\mb{n}}  \bracs{\cos \phi }^{\tr \mb{n} - \tr \mb{l}} \sqrt{\prod_{m=1}^M {n_m  \choose l_m}} \keti{\chi_\mb{n}}\kets{\mb{n} - \mb{l}} \\
&= \bracs{\sin \phi}^{\tr \mb{l}}\sum_{\mb{n} \geq \mb{0}} \sqrt{p_{\mb{n} + \mb{l}}} \bracs{\cos \phi }^{\tr \mb{n}} \sqrt{\prod_{m=1}^M {n_m  + l_m \choose l_m}} \keti{\chi_{\mb{n}+\mb{l}}} \kets{\mb{n}} \\
&\equiv \sum_{\mb{n} \geq \mb{0}} \sqrt{r_\mb{n}^{(\mb{l})}} \keti{\xi^{(\mb{l})}_\mb{n}} \kets{\mb{n}} \\
&\equiv \sqrt{q_\mb{l}} \ket{\psi_\mb{l}}
\end{aligned}
\end{equation}
an NDS state with squared norm $q_\mb{l} = \braket{\psi_\mb{l}}{\psi_\mb{l}}$. Similarly,
\begin{equation} \label{r'defined}
\begin{aligned}
\varket{\psi'_\mb{l}} &=
\bracs{\sin \phi'}^{\tr \mb{l}} \sum_{\mb{n} \geq \mb{l}} \sqrt{p_\mb{n}}  \bracs{\cos \phi' }^{\tr \mb{n} - \tr \mb{l}} \sqrt{\prod_{m=1}^M {n_m  \choose l_m}} \keti{\chi_\mb{n}}\kets{\mb{n} - \mb{l}} \\
&= \bracs{\sin \phi'}^{\tr \mb{l}}\sum_{\mb{n} \geq \mb{0}} \sqrt{p_{\mb{n} + \mb{l}}} \bracs{\cos \phi' }^{\tr \mb{n}} \sqrt{\prod_{m=1}^M {n_m  + l_m \choose l_m}} \keti{\chi_{\mb{n}+\mb{l}}} \kets{\mb{n}} \\
&\equiv \sum_{\mb{n} \geq \mb{0}} \sqrt{{r'}_\mb{n}^{(\mb{l})}} \keti{\xi^{(\mb{l})}_\mb{n}} \kets{\mb{n}} \\
&\equiv \sqrt{q'_\mb{l}} \ket{\psi'_\mb{l}}
\end{aligned}
\end{equation}
an NDS state with squared norm $q'_\mb{l} = \braket{\psi'_\mb{l}}{\psi'_\mb{l}}$. 

Since 
\begin{equation}
\begin{aligned}
\rho &= \pars{\mr{id}_A \otimes \cl{A}_{G}^{\otimes M}}\pars{\sigma}, \\
\rho' &= \pars{\mr{id}_A \otimes \cl{A}_{G'}^{\otimes M}}\pars{\sigma'},
\end{aligned}
\end{equation}
we can use the strong concavity of fidelity \cite{NC00} to write
\begin{align} \label{concavitybound}
F\pars{\rho,\rho'} &\geq \sum_{\mb{l} \geq \mb{0}} \sqrt{q_\mb{l} \,q'_\mb{l}}\; F\pars{\mr{id}_A \otimes \cl{A}_G^{\otimes M}\pars{\Psi_\mb{l}}, \mr{id}_A \otimes \cl{A}_{G'}^{\otimes M}\pars{\Psi'_\mb{l}}}.
\end{align}
We can now apply Eq.~\eqref{ampoutputfidelity} from Proposition \ref{prop:ampoutputfidelity} to write the summand above as
$\sum_{n=0}^\infty f_n^{(\mb{l})}\nu^{n+M}$ where 
\begin{align}
f_n^{(\mb{l})} = \sum_{\mb{n}: \tr \mb{n} = n} \sqrt{{r}_\mb{n}^{(\mb{l})} \,{r'}_\mb{n}^{(\mb{l})}}.
\end{align}
The right-hand side of \eqref{concavitybound} then becomes
\begin{align}
F\pars{\rho,\rho'} &\geq \sum_{\mb{l} \geq \mb{0}} \sum_{n=0}^\infty\nu^{n+M} \sum_{\mb{n}: \tr \mb{n} = n} \sqrt{{r}_\mb{n}^{(\mb{l})} \,{r'}_\mb{n}^{(\mb{l})}} \\
&=  \nu^M \sum_{\mb{l} \geq \mb{0}} \sum_{\mb{n} \geq \mb{0}} p_{\mb{n}+\mb{l}} \bracs{\sin \phi\, \sin \phi'}^{\tr \mb{l}} \bracs{\nu \,\cos \phi\, \cos \phi'}^{\tr \mb{n}} \pars{\prod_{m=1}^M {n_m  + l_m \choose l_m}}
\end{align}
Setting $\mb{k} = \mb{n} + \mb{l}$, the sum can be rewritten as
\begin{align}
F\pars{\rho,\rho'} &\geq \nu^M \sum_{\mb{k} \geq \mb{0}} p_{\mb{k}} \bracs{\nu \cos\phi \cos\phi'}^{\tr \mb{k}} \sum_{\mb{l} \leq \mb{k}} \bracs{\nu^{-1} \tan \phi \tan \phi'}^{\tr \mb{l}} \pars{\prod_{m=1}^M {k_m \choose l_m}}\\
&=  \nu^M \sum_{\mb{k} \geq \mb{0}} p_{\mb{k}} \bracs{\nu \cos\phi \cos \phi' + \sin\phi \sin \phi'}^{\tr \mb{k}}\\
&= \nu^M \sum_{n=0}^\infty p_{n} \bracs{\nu \sqrt{\kappa \kappa'} + \sqrt{(1-\kappa)(1-\kappa')}}^n.
\end{align}
\end{proof}
\end{thm}
\noindent Eq.~(61) corresponds to Eq.~(9) of the main text.
This generalizes a similar result derived for quantum-limited loss channels \cite{Nai11} that can be recovered by setting $G'=G=1$. A bound expressed only in terms of the total probe energy $\cl{N}_S = \sum_{n=0}^\infty n p_n$ can be obtained using Jensen's inequality.

Returning to Alice's target detection problem, we set $\kappa = 0, G = N_B+1,\kappa' = \eta/G', G' = (1-\eta) N_B + 1$ to get the lower bound
\begin{align}
F\pars{\rho_0,\rho_1} \geq \nu^M\sum_{n=0}^\infty p_n \pars{1- \frac{\kappa}{(1-\kappa)N_B+1}}^{n/2},
\end{align}
which in turn implies via the results of \cite{FvdG99} that
\begin{align}
P_{\frke}^A\geq \frac{1 - \sqrt{1 - F^2\pars{\rho_0,\rho_1} }}{2}, 
\end{align}
which leads to {Eq.~(10) of the main text}.

\section{Appendix C: $\epsilon$-covertness}

\subsection{General Definition}
As discussed in the main text (cf. Eq.~(6) therein), Willie faces the hypothesis test
\begin{equation} \label{Willietest}
\begin{aligned}
\mathrm{H}_0': \; \sigma_0 &= {\rho_\tsf{th}\pars{N_B}}^{\otimes M},\\
\mathrm{H}_1': \; \sigma_1 &=  \cl{L}_{1-\eta, N_B}^{\otimes M}\pars{\Tr_I \Psi}.
\end{aligned}
\end{equation}
to decide whether Alice has sent a probe ($\mr{H}_1'$) or not ($\mr{H}_0'$). Our definition of $\epsilon$-covertness (Eq.~(2) of the main text) assumes that the prior probabilities $\lambda_0$ and $\lambda_1$ of $\mr{H}_0'$ and $\mr{H}_1'$ are equal. More generally, Willie may have some prior information that leads to unequal $\lambda_0$ and $\lambda_1$. We then have $P_\frke^W \leq \min(\lambda_0,\lambda_1),$ which is saturated by Willie simply assuming the more likely alternative. As such, it makes sense to  define $\epsilon$-covertness as
\begin{align} \label{genecovertness}
P_\frke^W \geq \min\pars{\lambda_0,\lambda_1}- \epsilon.
\end{align}

\subsection{Necessary condition for $\epsilon$-covertness}
In this section, we derive a necessary condition for $\epsilon$-covertness. The condition applies to any $M$-signal-mode probe and is expressed in terms of the photon number probability mass function (pmf) of the \emph{total} photon number in the state $\sigma_1$ {of Eq.~(6) corresponding to Willie's alternative hypothesis $\mr{H}_1'$}. 

Willie's null hypothesis state is
\begin{align}
\sigma_0 &= \tau\pars{N_B}^{\otimes M} = \prod_{m=1}^M \bracs{\sum_{n_m=0}^\infty \frac{N_B^{n_m}}{\pars{N_B+1}^{n_m +1}}\ket{n_m}\bra{n_m}},\\
&= \sum_{\mb{n} \geq \mb{0}}  \frac{N_B^n}{\pars{N_B+1}^{n+M}} \ket{\mb{n}}\bra{\mb{n}}, 
\end{align}
where $\ket{\mb{n}}_W = \ket{n_1}\otimes\cdots\otimes\ket{n_M}$ are $M$-mode number states of Willie's modes and we have defined the total photon number $n := \tr\mb{n} = \sum_{m=1}^M n_m$.
A general  alternative-hypothesis state can be written as
\begin{align}
\sigma_1=\sum_{\mb{n},\mb{n}'\geq \mb{0}} q_{\textbf{n},\textbf{n}'}\ket{\textbf{n}}\bra{\textbf{n}'}.
\end{align}
The $\epsilon$-covertness condition Eq.~\eqref{genecovertness} reads
\begin{align}
\min\pars{\lambda_0,\lambda_1}- \epsilon \leq P_\frke^W = \frac{1}{2} - \frac{\norm{\lambda_0\sigma_0 - \lambda_1\sigma_1}_1}{2} \leq \sqrt{\lambda_0\,\lambda_1}\,{F\pars{\sigma_0,\sigma_1}}, \label{fvdg}
\end{align}
where $F\pars{\sigma_0,\sigma_1} = \Tr \sqrt{\sqrt{\sigma_0} \sigma_1 \sqrt{\sigma_0}}$ is the fidelity between $\sigma_0$ and $\sigma_1$ and the last inequality is the generalization  of the Fuchs-van de Graaf inequality \cite{FvdG99} to unequal priors \cite{Aud14}.

Let $\cl{M}$ be the channel corresponding to a (non-destructive) measurement of photon number in each of Willie's modes, i.e, $\cl{M}\sigma = \sum_{\mb{n} \geq \mb{0}} \bra{\mb{n}}\sigma \ket{\mb{n}} \ket{\mb{n}}\bra{\mb{n}}$ for any input state $\sigma$.
Using the data processing inequality for the fidelity, we can write
\begin{align}
F\pars{\sigma_0,\sigma_1} &\leq F\pars{\cl{M}\sigma_0,\cl{M}\sigma_1} \\
&= F\pars{\sigma_0,\cl{M}\sigma_1}\\
&= F\pars{\sigma_0, \sum_{\mb{n}} q_{\mb{n}} \ket{\mb{n}}\bra{\mb{n}}}\\
&= \sum_{n=0}^\infty \sqrt{\frac{N_B^n}{\pars{N_B+1}^{n+M}}} \pars{\sum_{\mb{n}: \tr \mb{n}=n}\sqrt{ q_{\mb{n}}}}, \label{covbnd1}
\end{align}
where we have set $q_{\mb{n}} := q_{\mb{n},\mb{n}}$. This bound is complicated in that it is expressed in terms of the photon number pmf of all of Willie's modes. To get a bound only in terms of the pmf $q_n = \sum_{\bn: n_1+\cdots+n_M=n} q_\bn$ of the total photon number, note that
\begin{align}    
 \sum_{\mb{n}: \tr \mb{n}=n}\sqrt{ q_{\mb{n}}} &\leq\sqrt{q_n}\sqrt{\binom{n+M-1}{n}},
\end{align}
where the binomial coefficient is the number of terms in the sum on the left-hand side. This inequality is saturated if $\{q_\bn\}$ is a product thermal distribution of any brightness. Substituting the above inequality into Eq.~\eqref{covbnd1} and using Eq.~\eqref{fvdg}, we get the necessary condition 
\begin{align} \label{genecovcondition}
    \sum_{n=0}^\infty\sqrt{q_n}\sqrt{\binom{n+M-1}{n}\frac{N_B^n}{(N_B+1)^{n+M}}}\geq \frac{\min\pars{\lambda_0,\lambda_1}-\epsilon}{\sqrt{\lambda_0\,\lambda_1}}.
\end{align}
Note that the effect of the prior probabilities is confined to the right-hand side and is insensitive to our later manipulations of the left-hand side. To avoid notational clutter, we have assumed in the main text that Willie's prior probabilities are equal, so that the necessary condition 
 for $\epsilon$-covertness becomes Eq.~(11) of the main text.

 \section{Appendix D: Range of allowed energies for an $\epsilon-$covert probe}
In this section, we obtain upper and lower limits on Alice's allowed signal energy beyond which $\epsilon$-covertness cannot hold. As discussed in the main text, this can be done by solving the problem:
\begin{align}
&\textbf{Extremise }f(q_1,q_2,\cdots)=\sum_{n=0}^\infty nq_n\nonumber\\
&\textbf{subject to }\nonumber\\
&\quad g(q_1,q_2,\cdots)=1-2\epsilon-\sum_{n=0}^\infty\sqrt{\binom{n+M-1}{n}\frac{N_B^n}{(N_B+1)^{n+M}}q_n}\leq 0\nonumber\\
&\quad h(q_1,q_2,\cdots)=\sum_{n=0}^\infty q_n-1=0\nonumber
\end{align}
where $g(q_1,q_2,\cdots)$ is the covertness constraint (Eq. (11) of the main text) and  $\{q_n\}$ is the pmf of Willie's total photon number under $\mathrm{H}_1'$. The Karush-Kuhn-Tucker (KKT) necessary conditions are as follows:\\
\textbf{Stationarity}\\
To find the point $\vec{q}^*$ that minimises $f(q_1,q_2,\cdots)$ across all discrete probability distributions $\vec{q}$, the required condition is
\begin{align}
\frac{\partial}{\partial\vec{q}}\left.\left[f+\lambda_1(g)+\lambda_2(h)\right]\right|_{\vec{q}=\vec{q}^*}=0,
\end{align}
which gives the system of equations for $n=0,1,2,\cdots$ such that 
\begin{align}
&\frac{\partial}{\partial q_n}\left.\left[-\sum_{n=0}^\infty nq_n+\lambda_1\left(1-2\epsilon-\sum_{n=0}^\infty\sqrt{\binom{n+M-1}{n}\frac{N_B^n}{(N_B+1)^{n+M}}q_n}\right)+\lambda_2\left(\sum_{n=0}^\infty q_n-1\right)\right]\right|_{q_n=q_n^*}=0\nonumber\\
&-n-\lambda_1\sqrt{\binom{n+M-1}{n}\frac{N_B^n}{(N_B+1)^{n+M}}}\left(\frac{1}{2\sqrt{q_n^*}}\right)+\lambda_2=0,\nonumber\\
&q_n^*=\frac{\lambda_1^2}{4(k-\lambda_2)^2}\left[\binom{n+M-1}{n}\frac{N_B^n}{(N_B+1)^{n+M}}\right]\label{eq:lagrange1}
\end{align}
where $\lambda_1$ and $\lambda_2$ are the KKT multipliers.\\
\textbf{Primal feasibility}
\begin{align}
&\sum_{n=0}^\infty q_n^*-1=0\hspace{250pt}\\
&1 - 2\epsilon-\sum_{n=0}^\infty\sqrt{\binom{n+M-1}{n}\frac{N_B^n}{(N_B+1)^{n+M}}q_n^*}\leq 0
\end{align}
\textbf{Dual feasibility}
\begin{align}
&\lambda_1\geq 0\hspace{290pt}
\end{align}
\textbf{Complementary slackness}
\begin{align}
\lambda_1\left(1-2\epsilon-\sum_{n=0}^\infty\sqrt{\binom{n+M-1}{n}\frac{N_B^n}{(N_B+1)^{n+M}}q_n^*}\right)=0\hspace{190pt}
\end{align}
From the stationarity condition, it can be concluded that $\lambda_1\neq 0$ or the normalization condition for $\{q_n\}$ will be violated. Hence, by applying this condition to the complementary slackness condition, it can be deduced that $\sum_{n=0}^\infty\sqrt{\binom{n+M-1}{n}\frac{N_B^n}{(N_B+1)^{n+M}}q_n^*}=1-2\epsilon$. In this case, the optimal solution is on the boundary of the constraint $g(q_1,q_2,\cdots)$. Substituting Eq. (\ref{eq:lagrange1}) into both constraints and confining the upper summation limit to a finite value yields the following simultaneous equations
\begin{align}
&\sum_{n=0}^d\mathcal{N}\frac{\lambda_1^2}{4(n-\lambda_2)^2}\left[\binom{n+M-1}{n}\frac{N_B^n}{(N_B+1)^{n+M}}\right]=1\nonumber\\
&\sum_{n=0}^d \mathcal{N}\left[\binom{n+M-1}{n}\frac{N_B^n}{(N_B+1)^{n+M}}\right]\abs{\frac{\lambda_1}{2(n-\lambda_2)}}=1-2\epsilon
\end{align}
where $d$ is the upper summation limit, $\mathcal{N}$ is the re-normalisation factor for finite summation. By selecting appropriate initial values for the KKT multipliers $\lambda_1$ and $\lambda_2$ in \textsc{Matlab}'s numerical solver, both the upper limit and lower limit of the signal strength intercepted by Wille are obtained. Given the relationship between Alice's allowable signal energies and Willie's intercepted energies $\mathcal{N}_S = (\sum_n nq_n -\eta N_B)/(1-\eta)$, the upper and lower allowable probe energies given $\epsilon$-covertness are obtained, as displayed in Fig.~2 of the main text.

\section{Appendix E: Photon-number probability generating functions (PGFs)}

In this section, we show how the pmf of the total photon number in  $W$ (Willie's modes) can be connected to the pmf of the total photon number in $S$ (the transmitter). To do so, we first review the notion of the photon number probability generating function and discuss its transformation under the pure-loss and quantum-limited amplifier channels, building on the single-mode treatment in ref.~\cite{Hau00enqom}, Ch.~9. 

\subsection{Three kinds of generating functions}
Let $\cl{H}$ be the Hilbert space of some fixed $M$-mode quantum system. Given a state $\rho \in \cl{S}\pars{\cl{H}}$, its \emph{photon-number probability generating function (pgf)} (henceforth `the pgf') is the function of the $M$-vector $\bxi = \pars{\xi_1,\ldots, \xi_M} \in \mathbb{R}^M$ defined as
\begin{align} \label{pgf}
\cl{P}\pars{\bxi} : = \sum_{\bn \geq \mb{0}} p_{\bn} \prod_{m=1}^M\xi_m^{n_m},
\end{align}
where $p_{\bn} = \bra{\bn}\rho\ket{\bn}$ is the photon number pmf on the $M$ modes of the system. Eq.~\eqref{pgf} is the straightforward multimode generalization of the single-mode definition \cite{Hau00enqom}. Because of the normalization $\sum_{\bn} p_{\bn} =1$, the pgf is seen to exist at least for all $\bxi$ such that $\abs{\xi_m}\leq 1$ for all $m$ between $1$ and $M$. 
The photon-number pmf can be recovered from it via
\begin{align}
p_{\bn} = \frac{\bracs{\partial_1^{n_1} \cdots \partial_M^{n_m} \cl{P}\pars{\bxi}}\Big\vert_{\bxi = \mb{0}}}{n_1!\cdots n_M!},
\end{align}
where $\partial_m \equiv \partial/\partial \xi_m$. Moreover, it is easily checked that the pgf of the \emph{total} photon number $p_n = \sum_{\bn: \tr \bn = n} p_{\bn}$  in $S$ can be obtained from $\cl{P}\pars{\bxi}$ as
\begin{align} \label{tpnpgf}
\cl{P}\pars{\xi} := \sum_{n=0}^\infty p_n \xi^n = \cl{P}\pars{\xi, \ldots, \xi}; \qquad \xi \in \mathbb{R}.
\end{align}
To avoid proliferation of notation, we use the same symbol for the multimode pgf and that of the total photon number, letting the argument specify which is meant.\newline

Let $\br = \pars{r_1,\ldots, r_M}$ be a vector of nonnegative integers. The $\br$-th \emph{falling factorial moment} of $\rho$ is defined -- generalizing the single-mode definition (cf. Sec.~9.1 of ref.~\cite{Hau00enqom}) -- as
\begin{align}
F_{\br}: = \Tr \rho \bracs{\prod_{m=1}^M {\ahat_m}^{\dag r_m}{\ahat_m}^{r_m}},
\end{align}
where $\{\ahat_m\}_{m=1}^M$ are the annihilation operators of the $M$ modes of the system. The \emph{falling-factorial moment generating function (falling-factorial mgf)} is then defined as
\begin{align} \label{ffmgf}
\cl{F}\pars{\bxi} : = \sum_{\br \geq \mb{0}} F_{\br} \frac{\xi_1^{r_1}}{r_1!}\cdots\frac{\xi_M^{r_M}}{r_M!}.
\end{align}
We can generalize the derivation of the single-mode correspondence given in Eq.~(9.4) of ref.~\cite{Hau00enqom} to obtain the relation 
\begin{align} \label{PtoF}
\cl{F}\pars{\bxi} = \cl{P}\pars{1+\xi_1,\ldots, 1 + \xi_M},
\end{align}
between the falling-factorial mgf and the pgf for an $M$-mode state.\newline

Similarly, for any $\br = \pars{r_1,\ldots, r_M}$, the $\br$-th \emph{rising factorial moment} of $\rho$ is defined  as
\begin{align}
R_{\br}: = \Tr \rho \bracs{\prod_{m=1}^M {\ahat_m}^{r_m}{\ahat_m}^{\dag r_m}}.
\end{align}
The \emph{rising-factorial moment generating function (rising-factorial mgf)} is then defined as
\begin{align} \label{rfmgf}
\cl{R}\pars{\bxi} : = \sum_{\br \geq \mb{0}} R_{\br} \frac{\xi_1^{r_1}}{r_1!}\cdots\frac{\xi_M^{r_M}}{r_M!}.
\end{align}
This object is related to the pgf of $\rho$ via
\begin{align} \label{PtoR}
\cl{R}\pars{\bxi} &=\bracs{\prod_{m=1}^M\pars{\frac{1}{1-\xi_m}}} \cl{P}\pars{\frac{1}{1-\xi_1},\cdots,\frac{1}{1-\xi_M}},
\end{align}
which generalizes the single-mode relation given in Eq.~(9.6) of ref.~\cite{Hau00enqom}. 
Thus, any of these three functions can be derived from any other.\newline

\subsection{PGF at the output of a thermal loss channel}

The falling-factorial and rising-factorial mgfs are useful because they transform in a simple manner under the action of pure-loss and quantum-limited amplifier channels respectively. For a single-mode state $\rho$, let $\rho_{\tsf{out}} = \cl{L}_\kappa \rho$,  the falling-factorial mgf $\cl{F}_{\tsf{out}}\pars{\xi}$ of $\rho_{\tsf{out}}$ is related to the falling-factorial mgf $\cl{F}\pars{\xi}$ of $\rho$ as:
\begin{align} \label{1modeloss}
\cl{F}_{\tsf{out}}\pars{\xi} = \cl{F}\pars{\kappa \xi}.
\end{align}
Similarly, if  $\rho_{\tsf{out}} = \cl{A}_G \rho$, the rising-factorial mgf $\cl{R}_{\tsf{in}}\pars{\xi}$ of $\rho_{\tsf{out}}$ is related to the rising-factorial mgf $\cl{R}\pars{\xi}$ of $\rho$ as:
\begin{align} \label{1modegain}
\cl{R}_{\tsf{out}}\pars{\xi} = \cl{R}\pars{G \xi};
\end{align}
see Eqs.~(9.27) and (9.36) in ref.~\cite{Hau00enqom}.

If $\rho$ is an $M$-mode state and $\rho_{\tsf{out}} = \cl{L}_\kappa^{\otimes M} \rho$, we can generalize the derivation of Eq.~\eqref{1modeloss} to show that the $M$-mode falling-factorial of Eq.~\eqref{ffmgf} transforms as
\begin{align} \label{Mmodeloss}
\cl{F}_{\tsf{out}}\pars{\bxi} = \cl{F}\pars{\kappa \bxi}.
\end{align}
On the other hand, if $\rho_{\tsf{out}} = \cl{A}_G^{\otimes M} \rho$,  the $M$-mode rising-factorial mgf of Eq.~\eqref{rfmgf} transforms as
\begin{align} \label{Mmodegain}
\cl{R}_{\tsf{out}}\pars{\bxi} = \cl{R}\pars{G \bxi}.
\end{align}
\newline
Now suppose that $\cl{C}$ is a product thermal loss channel that has the decomposition
\begin{align}
\cl{C} = \cl{A}_G^{\otimes M} \circ \cl{L}_{\kappa}^{\otimes M}.
\end{align}
By using Eqs.~\eqref{Mmodeloss}-\eqref{Mmodegain} together with the relations of Eqs.~\eqref{PtoF}-\eqref{PtoR} relating the mgfs to the pgf, we can show that the pgf $P_{\tsf{out}}\pars{\bxi}$ of the output state $\rho_{\tsf{out}} = \cl{C}\rho$ is related to the pgf $P\pars{\bxi}$ of $\rho$ via
\begin{align} \label{outputpgf}
\cl{P}_{\tsf{out}}\pars{\bxi} &= \bracs{\prod_{m=1}^M\frac{1}{G-\xi_m(G-1)}}\,\cl{P}\pars{1- \kappa + \frac{\kappa\xi_1}{G-\xi_1(G-1)}, \ldots, 1- \kappa + \frac{\kappa\xi_M}{G-\xi_M(G-1)}}.
\end{align}
In particular, the pgf $\cl{P}_{\tsf{out}}(\xi)$ of the total photon number in the output of the thermal loss channel can be related to the pgf $P\pars{\xi}$ of the total photon number in the input via Eq.~\eqref{tpnpgf} as:
\begin{align} \label{pgftrans}
\cl{P}_{\tsf{out}}(\xi) &= \bracs{\frac{1}{G-\xi(G-1)}}^M \,\cl{P}\pars{1- \kappa + \frac{\kappa\xi}{G-\xi(G-1)}}.
\end{align}

\section{Appendix F: Lower bound on Alice's error probability under $\epsilon$-covertness}
In this section, we derive the lower bound of Eq.~(13) of the main text that must be satisfied under $\epsilon-$ covertness.
\subsection{Analytical bound}
We begin by rewriting the $\epsilon-$ covertness condition of Eq.~(11) of the main text (assuming equal priors for the moment) as
\begin{align}
    &\sum_{n=0}^\infty\sqrt{\binom{n+M-1}{n}\frac{N_B^n}{(N_B+1)^{n+M}}x^{-n}x^nq_n}\geq 1-2\epsilon,
    \end{align}
where $x$ will be specified later.    
 Using the Cauchy-Schwartz inequality, the expression can be bounded as a function of the input probability generating function
    \begin{align}
    &\sum_{n=0}^\infty\sqrt{\binom{n+M-1}{n}\frac{N_B^n}{(N_B+1)^{n+M}}x^{-n}x^nq_n}\geq 1-2\epsilon\nonumber\\
    &\left(\sum_{n=0}^\infty\binom{n+M-1}{n}\frac{N_B^n}{(N_B+1)^{n+M}}x^{-n}\right)^{1/2}\left(\sum_{n=0}^\infty q_n x^n\right)^{1/2}\geq 1-2\epsilon\nonumber\\
    &\left(\frac{1}{N_B+1-\frac{N_B}{x}}\right)^M\cl{P}_W(x)\geq(1-2\epsilon)^2,
\end{align}
where $\cl{P}_W(x)$ is the probability generating function of $\sigma_1$. For the first sum to converge, we must have $N_B/\abs{x}\leq N_B+1$. Using Eq.~\eqref{pgftrans} to express the above inequality in terms of the pgf $\cl{P}_S$ of the probe gives \begin{align}
    &\cl{P}_S\left(1+\frac{(x-1)(1-\eta)}{\eta N_B(1-x)+1}\right)\geq (1-2\epsilon)^2\pars{N_B+1 - \frac{N_B}{x}}^M\bracs{\eta N_B (1-x)+1}^M.
\end{align}
The  lower bound (Eq.~(10) of the main text) on $P_\frke^A$ for QI target detection is a function of the fidelity between $\rho_0$ and $\rho_1$, namely $P_{\frke}^A\geq 1/2 - \sqrt{1- F^2(\rho_0,\rho_1)}/2$, where the fidelity is lower-bounded in terms of the probe pgf as follows:
\begin{align}
  F(\rho_0,\rho_1) &\geq \nu^M\sum_{n=0}^\infty p_n\left[1 - \frac{\eta}{(1-\eta)N_B+1}\right]^{n/2}  \nonumber\\
  &=\nu^M \cl{P}_S\left(\left(1-\frac{\eta}{(1-\eta)N_B+1}\right)^{1/2}\right)
\end{align}
Equating  the arguments of the pgf in Eqns.~(87) and (88), we can solve for the independent variable $x$:
\begin{align}
    &1+\frac{(x-1)(1-\eta)}{\eta N_B(1-x) + 1}=\left[1 - \frac{\eta}{(1-\eta)N_B + 1}\right]^{1/2}\\
    \Rightarrow & x=1 - \frac{\Theta}{\eta[1+N_B(1-\Theta)]},\qquad \Theta=\frac{\sqrt{(1-\eta)(N_B+1)}}{\sqrt{1 + (1-\eta)N_B}}. \label{x}
\end{align}
Applying the convergence condition on $x$  yields:
\begin{align}
    \frac{N_B}{N_B+1}\leq 1 - \frac{\Theta}{\eta[1+N_B(1-\Theta)]}\leq 1
\end{align}
In the limit $\eta \ll 1$, expanding $x$ up to first order of $\eta$ gives
\begin{align}
    x = 1 - \frac{\eta}{2(N_B+1)}+\mathcal{O}(\eta^2),
\end{align}
it can be deduced that in this regime, the convergence condition is always true for finite values of $N_B$. Numerically, the above condition is satisfied for $\eta\leq 0.4$. Thus, the lower bound of the fidelity between Alice's received states can be expressed as 
\begin{align} \label{ecovfidlb}
    F(\rho_0,\rho_1)\geq (1-2\epsilon)^2\nu^M\left(N_B+1-\frac{N_B}{x}\right)^M[\eta N_B(1-x) + 1]^M.
\end{align}
Applying the Fuchs-van de Graaf inequality gives rise to Eq.~(13) of the main text with
\begin{align} 
f\pars{\eta,N_B,M} = \nu^{2M}\left(N_B+1-\frac{N_B}{x}\right)^{2M}[\eta N_B(1-x) + 1]^{2M},
\end{align}
with $x$ given by Eq.~\eqref{x}.
For the case of \emph{unequal priors} $\lambda_0 \neq \lambda_1$ for Willie's hypothesis test, we start with the covertness condition of Eq.~\eqref{genecovcondition}. It can be verified that the result is the same as Eq.~\eqref{ecovfidlb} with $1-2\epsilon$ replaced by $\frac{\min\pars{\lambda_0,\lambda_1}-\epsilon}{\sqrt{\lambda_0\,\lambda_1}}$.

\subsection{Numerical evaluation}
To check the tightness of the above bound, we also numerically optimized Eq.~(11) of the main text under the covertness constraint. As mentioned above, the fidelity between the states received by Alice can be rewritten as:
\begin{align}
     F(\rho_0,\rho_1) &\geq \nu^M \sum_{n=0}^\infty p_n\left[1-\frac{\eta}{(1-\eta)N_B+1}\right]^{n/2}\nonumber\\
    &=\nu^M\mathcal{P}_S\left(\xi\right)\nonumber\\
    &=\nu^M\left[\frac{G_W\eta_W}{\xi(G_W - 1) - G_W(1 - \eta_W) + 1}\right]^M\mathcal{P}_W\left(1 - \frac{1 - \xi}{G_W(\eta_w + \xi - 1) - \xi + 1}\right)\nonumber\\
    &=\nu^M\left[\frac{G_W\eta_W}{x(G_W - 1) - G_W(1 - \eta_W) + 1}\right]^M\sum_{n=0}^\infty q_n x^n,
\end{align}
where $G_W = \eta N_B+1$, $\eta_W = (1-\eta)/G_W$, $\xi = \sqrt{ 1 - \frac{\eta}{(1-\eta)N_B + 1}}$ and $x = 1 - \frac{1-\xi}{G_W(\eta_W + \xi - 1) - \xi + 1}$. The problem statement is hence\\
\textbf{Problem statement:}
\begin{align}
\textbf{Minimise: }&f(q_1,q_2,\cdots)=\nu^M\left[\frac{G_W\eta_W}{x(G_W - 1) - G_W(1 - \eta_W) + 1}\right]^M\sum_{n=0}^\infty q_n x^n\nonumber\\
\textbf{Subject to: }&g(q_1,q_2,\cdots)=\sum_{n=0}^\infty\sqrt{\binom{n+M-1}{n}\frac{N_B^n}{(N_B+1)^{n+M}}q_n}\geq 1-2\epsilon\nonumber\\
&h(q_1,q_2,\cdots)=\sum_{n=0}^\infty q_n = 1,
\end{align}
where $g(q_1,q_2,\cdots)$ is the inequality constraint while $h(q_1,q_2,\cdots)$ is the equality constraint. The Karush-Kuhn-Tucker (KKT) necessary conditions for non-linear optimisation are as follows:\\
\textbf{Stationarity}\\
To find the point $\Vec{q}^*$ that minimises $f(q_1,q_2,\cdots)$ across all discrete probability distributions $\Vec{q}$, the required condition is
\begin{align}
    \frac{\partial}{\partial q_n}\left.\left[f + \lambda_1 g + \lambda_2 n\right]\right|_{q_n = q_n^*}
\end{align}
which gives the system of equation, $n = 0,1,2,\cdots$ such that
\begin{align}
    q_n^* = \frac{\lambda_1^2 p_n}{4(\nu^M\mu^Mx^n + \lambda_2)^2}
\end{align}
where $\lambda_1$ and $\lambda_2$ are the KKT multipliers, $p_n = \binom{n+M-1}{n}\frac{N_B^n}{(N_B+1)^{n+M}}$, and $\mu = \frac{G_W\eta_W}{x(G_W - 1) - G_W(1 - \eta_W) + 1}$.\\
\textbf{Primal feasibility}
\begin{align}
  &1-2\epsilon - \sum_{n=0}^\infty\sqrt{\binom{n+M-1}{n}\frac{N_B^n}{(N_B+1)^{n+M}}q_n}\leq 0 \\
  &\sum_{n=0}^\infty q_n - 1 = 0
\end{align}
\textbf{Dual feasibility}
\begin{align}
    \lambda_1\geq 0
\end{align}
\textbf{Complementary slackness}
\begin{align}
    \lambda_1\left(1-2\epsilon - \sum_{n=0}^\infty\sqrt{\binom{n+M-1}{n}\frac{N_B^n}{(N_B+1)^{n+M}}q_n}\right) = 0
\end{align}
From the stationarity condition, it can be concluded that $\lambda_1\neq 0$ or else the equality constraint is violated. Hence, by applying this condition to the complementary slackness condition, it can be deduced that 
\begin{align} \nonumber
\left(1-2\epsilon - \sum_{n=0}^\infty\sqrt{\binom{n+M-1}{n}\frac{N_B^n}{(N_B+1)^{n+M}}q_n}\right) = 0.
\end{align}
In this case, the optimal solution is on the boundary of the constraint $g(q_1,q_2,\cdots)$. Hence, expressing the two constraints as equality and confining the upper summation limit to a finite value,
\begin{align}
    &\sum_{n=0}^d\mathcal{N}\binom{n+M-1}{n}\frac{N_B^n}{(N_B+1)^{n+M}}\frac{\abs{\lambda_1}}{2\abs{\nu^M\mu^M x^n + \lambda_2}} = 1-2\epsilon\\
    &\sum_{n=0}^d\mathcal{N}\binom{n+M-1}{n}\frac{N_B^n}{(N_B+1)^{n+M}}\frac{\lambda_1^2}{4(\nu^M\mu^M x^n + \lambda_2)^2}=1
\end{align}
where $d$ is the upper summation limit, $\mathcal{N}$ is the re-normalisation factor for finite summation. Using numerical solver on \textsc{Matlab} with appropriate initial values of $\lambda_1$ and $\lambda_2$, the above simultaneous equations can be solved and the optimal target detection fidelity is obtained. The heat map of Fig.~1 compares the numerically optimized target detection fidelity and the analytical bound obtained in Equation (107) above. We see that, apart from some numerical artefacts appearing for some values of $\epsilon$ for $N_B=0.2$, the bound is in good agreement with the numerical minimum.

\begin{figure}
     \centering
     \begin{subfigure}[b]{0.3\textwidth}
         \centering
         \includegraphics[width=\textwidth]{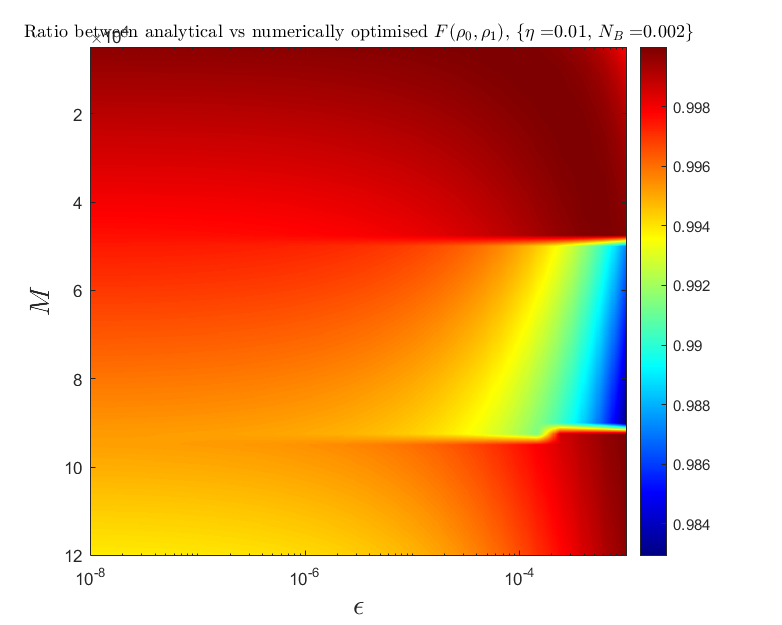}
         \caption{$N_B = 0.002$}
         \label{Nb=0.002}
     \end{subfigure}
     \hfill
     \begin{subfigure}[b]{0.3\textwidth}
         \centering
         \includegraphics[width=\textwidth]{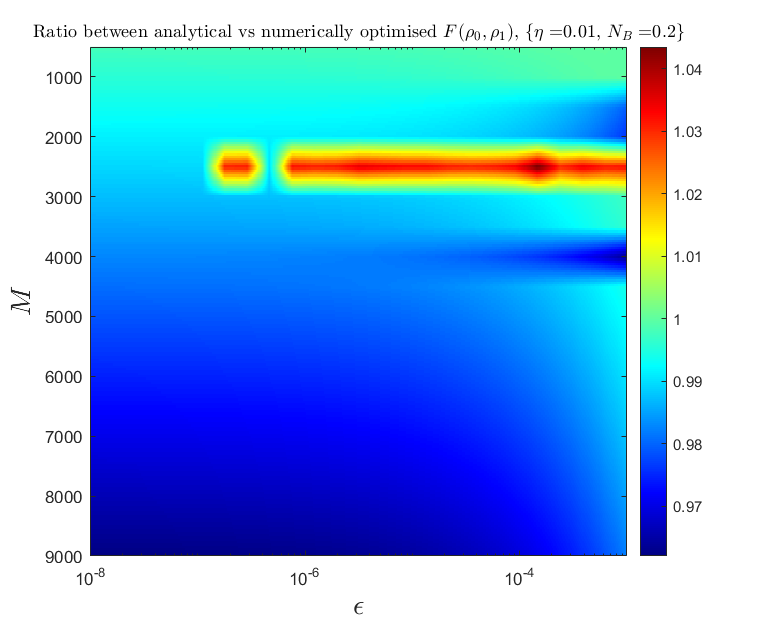}
         \caption{$N_B = 0.2$}
         \label{Nb=0.2}
     \end{subfigure}
     \hfill
     \begin{subfigure}[b]{0.3\textwidth}
         \centering
         \includegraphics[width=\textwidth]{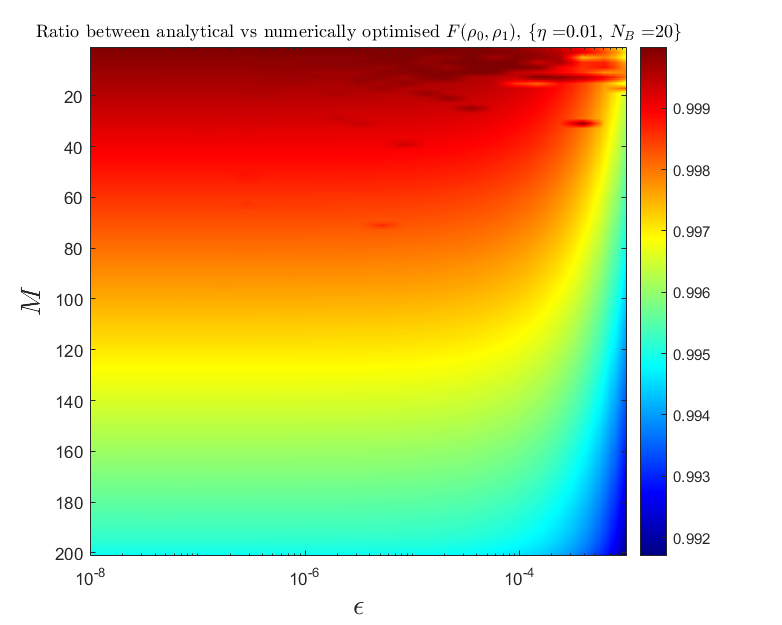}
         \caption{$N_B = 20$}
         \label{Nb=20}
     \end{subfigure}
        \caption{ Ratio of the analytical lower bound of Eq.~(107) to the numerically minimized fidelity under $\epsilon-$ covertness.}
        \label{analyticalvsnumerical}
\end{figure}

\subsection{Performance of TMSV and GCS probes under $\epsilon-$covertness}
When either TMSV or GCS states are used as probes by Alice, the states intercepted by Willie will be thermal states. To ensure that we use a brightness $N_S$ for these states that satisfies $P_\frke^W \geq 1/2 - \epsilon$, we evaluate the error probability and require that 
\begin{align}
    \frac{1}{2} - \frac{1}{4}\norm{\sigma_0-\sigma_1}_1&\geq \frac{1}{2}-\epsilon\nonumber\\
    \norm{\sigma_0-\sigma_1}_1&\leq 4\epsilon.
\end{align}
The trace norm between the two states intercepted by Willie can be calculated as
\begin{align}
    \norm{\sigma_0-\sigma_1}_1 &= \Tr\left[\sum_{\mb{n}\geq\mb{0}}\abs{\frac{N_B^n}{(N_B+1)^{n+M}}-\frac{N^n}{(N+1)^{n+M}}}\ket{\mb{n}}\bra{\mb{n}}\right]\nonumber\\
    &=\sum_{n=0}^\infty\binom{n+M-1}{n}\abs{\frac{N_B^n}{(N_B+1)^{n+M}}-\frac{N^n}{(N+1)^{n+M}}}\nonumber\\
    &=\sum_{n=0}^{n_t}\binom{n+M-1}{n}\frac{N_B^n}{(N_B+1)^{n+M}} -\sum_{n=0}^{n_t}\binom{n+M-1}{n}\frac{N^n}{(N+1)^{n+M}}\nonumber\\
    &\hspace{50pt}+\sum_{n=n_t+1}^\infty\binom{n+M-1}{n}\frac{N^n}{(N+1)^{n+M}}-\sum_{n=n_t+1}^\infty\binom{n+M-1}{n}\frac{N_B^n}{(N_B+1)^{n+M}},
\end{align}
where $n\coloneq\tr\mb{n}=\sum_{m=1}^Mn_m$, $N = (1-\eta)N_S + \eta N_B$, and $n_t = \left\lfloor \left.M\ln\left[\frac{N+1}{N_B+1}\right]\right/\ln\left[\frac{N(N_B+1)}{(N+1)N_B}\right]\right\rfloor$. Solving it numerically would require the evaluation of the binomial coefficient up to a very large value of $M$ and $n$, usually resulting in numerical overflow. As such, a trick is deployed where we instead calculate the logarithm of the binomial coefficient. A brief explanation of the method used to compute the first summation of the trace norm is discussed below:
\begin{align}
    \sum_{n=0}^{n_t}\binom{n+M-1}{n}\frac{N_B^n}{(N_B+1)^{n+M}}&=\sum_{n=0}^{n_t}e^{\ln\left[\binom{n+M-1}{n}\frac{N_B^n}{(N_B+1)^{n+M}}\right]}.
\end{align}
The exponent is evaluated as
\begin{align}
    \ln\left[\binom{n+M-1}{n}\frac{N_B^n}{(N_B+1)^{n+M}}\right]&=\ln\left[\frac{\Gamma(n+M)}{\Gamma(n+1)\Gamma(M)}\frac{N_B^n}{(N_B+1)^{n+M}}\right]\nonumber\\
    &=\ln\Gamma(n+M) - \ln\Gamma(n+1) - \ln\Gamma(M) + n\ln N_B - (n+M)\ln(N_B+1),
\end{align}
where the gamma function $\Gamma(x) = (x-1)!$ can be approximated to a high order of accuracy using Lanczos approximation. Deploying the same method to evaluate the summations for the other three terms and selecting a sufficiently large value for the upper limit of summation, we can thus compute the error probability for given values of $M, N_B$, and $N_S$. For fixed $M$ and $N_B$, we can vary $N_S$ to obtain the maximum $N_S$ such that TMSV or GCS probes of that brightness are guaranteed to be $\epsilon$-covert. Using the quantum Chernoff bound, we calculate the performance of TMSV and GCS states for the respective maximum $N_S$ as shown in Fig.~3 of the main paper.

\bibliography{covertrefsv3.bib}
\end{document}